\documentclass[twocolumn]{autart}    % Enable this line and disable the 
\usepackage{graphicx} % Required for inserting images
\usepackage{moreverb}
\usepackage{blindtext}
\usepackage[english]{babel}
\usepackage[utf8]{inputenc}
\usepackage{graphicx}
\usepackage{bbding}
\usepackage{amsmath}
\usepackage{amssymb}
\let\theoremstyle\relax
\usepackage{amsthm}
\usepackage{tikz} 
\usepackage{caption}
\usepackage{graphics}
\usepackage{graphicx}
\usepackage{tabularx}
\usepackage{array}

\usepackage{cite}
\usepackage{wrapfig}
\usepackage{mathrsfs}
\usepackage{url, soul}
\usepackage{color}
\newtheorem{theorem}{Theorem}[section]
\newtheorem{lemma}[theorem]{Lemma}
\newtheorem{definition}[theorem]{Definition}
\newtheorem{proposition}[theorem]{Proposition}
\newtheorem{assumption}[theorem]{Assumption}
\newtheorem{corollary}[theorem]{Corollary}
\newtheorem{remark}[theorem]{Remark}
\usepackage{graphicx}
\usepackage{epstopdf}
\usepackage{lipsum}
\usepackage{caption}
\usepackage{blindtext}
\usepackage{epstopdf}
\usepackage{mathtools}

\usepackage{subfigure}
\newcommand{\St}{\mathtt{S}}

\newcommand{\Pt}{\mathtt{P}}
\newcommand{\Ut}{\mathtt{U}}
\newcommand{\zs}{z_{\mathtt{S}}}
\newcommand{\zsne}{\zs^{\mathtt{NE}}}
\newcommand{\cp}{\mathtt{C}_{\mathtt{P}}}

\newcommand{\cmax}{\mathtt{C}_{\mathtt{\max}}}
\newcommand{\cmin}{\mathtt{C}_{\mathtt{\min}}}

\newcommand{\Ht}{\mathtt{H}}
\newcommand{\Lt}{\mathtt{L}}
\newcommand{\rhoh}{\rho_{\mathtt{H}}}
\newcommand{\rhol}{\rho_{\mathtt{L}}}

\newcommand{\Ststar}{\mathtt{S}^{\star}}
\newcommand{\Il}{\mathtt{I}_{\mathtt{L}}}

\newcommand{\Ilstar}{\mathtt{I}^{\star}_{\mathtt{L}}}

\newcommand{\Ih}{\mathtt{I}_{\mathtt{H}}}

\newcommand{\Ihstar}{\mathtt{I}^{\star}_{\mathtt{H}}}
\newcommand{\betah}{\beta_{\mathtt{H}}}
\newcommand{\hatbetah}{\hat{\beta}_{\mathtt{H}}}
\newcommand{\betal}{\beta_{\mathtt{L}}}
\newcommand{\hatbetal}{\hat{\beta}_{\mathtt{L}}}
\newcommand{\gammah}{\gamma_{\mathtt{H}}}
\newcommand{\gammal}{\gamma_{\mathtt{L}}}
\newcommand{\qhl}{q_\mathtt{HL}}
\newcommand{\qlh}{q_\mathtt{LH}}

\newcommand\um[1]{{#1}}

\usepackage{enumerate}

\begin{document}

\begin{frontmatter}

\title{\um{Bi-Virus SIS Epidemic Propagation under Mutation and Game-theoretic Protection Adoption}\thanksref{footnoteinfo}} % Title, preferably not more 

\thanks[footnoteinfo]{\um{A preliminary version of this work is accepted, and will appear in the \emph{Proceedings of the IEEE Conference on Decision and Control (CDC), 2025.} The conference version includes the characterization of stationary Nash equilibrium under the special case of uni-directional mutation, whereas, this work extends the analysis to the general case of bi-directional mutation.}}

\author[Kharagpur]{Urmee Maitra}\ead{urmeemaitra93@kgpian.iitkgp.ac.in},    % Add the 
\author[Kharagpur]{Ashish R. Hota}\ead{ahota@ee.iitkgp.ac.in},               % e-mail address 
\author[East Lansing]{Vaibhav Srivastava}\ead{vaibhav@msu.edu}  % (ead) as shown

\address[Kharagpur]{Department of Electrical Engineering, IIT Kharagpur, Kharagpur, West Bengal, India, 721302}  % Please supply                                              
\address[East Lansing]{Department of Electrical and Computer Engineering, Michigan State University, East Lansing, MI 48824}        % here.

\begin{abstract}                          % Abstract of not more than 200 words.
\um{We study a bi-virus susceptible-infected-susceptible (SIS) epidemic model in which individuals are either susceptible or infected with one of two virus strains, and consider mutation-driven transitions between strains. The general case of bi-directional mutation is first analyzed, where we characterize the disease-free equilibrium and establish its global asymptotic stability, as well as the existence, uniqueness, and stability of an endemic equilibrium. We then present a game-theoretic framework where susceptible individuals strategically choose whether to adopt protection or remain unprotected, to maximize their instantaneous payoffs. We derive Nash strategies under bi-directional mutation, and subsequently consider the special case of uni-directional mutation. In the latter case, we show that coexistence of both strains is impossible when mutation occurs from the strain with lower reproduction number and transmission rate to the other strain. Furthermore, we fully characterize the stationary Nash equilibrium (SNE) in the setting permitting coexistence, and examine how mutation rates influence protection adoption and infection prevalence at the SNE. Numerical simulations corroborate the analytical results, demonstrating that infection levels decrease monotonically with higher protection adoption, and highlight the impact of mutation rates and protection cost on infection state trajectories.}
\end{abstract}

\end{frontmatter}

\section{Introduction}

\um{Containing the spread of infectious diseases is a major challenge for policy-makers. As witnessed during the recent COVID-19 pandemic, the epidemic can spread over waves, potentially driven by the emergence of multiple viral strains through mutations \cite{nowak1991evolution,chavda2022delta} or the selfish response of individuals towards protection adoption \cite{paarporn2023sis,satapathi2022epidemic}. While early studies primarily investigated the dynamics of a single virus, recent works have extended these models to accommodate the coexistence of multiple viruses \cite{prakash2012winner,pare2021multi,janson2020networked,watkins2015optimal,she2022networked,zhang2022networked,santos2015sufficient}.

In addition to intrinsic factors such as transmission and recovery rates, individual behavior regarding protection adoption plays a crucial role in epidemic dynamics. This motivated researchers to view the problem of epidemic propagation coupled with human behavior through the prism of game theory. Prior research has explored the influence of human decision-making on epidemic spread using game-theoretic frameworks \cite{reluga2010game,paarporn2023sis,martins2023epidemic,hota2023learning,maitra2023sis,10750281,parino2024optimal}. We refer the readers to \cite{huang2022game} for a comprehensive review. However, despite extensive literature on epidemic games, few studies have examined the combined effect of game-theoretic decision-making and viral mutation on the coexistence and survival of multiple strains. Our work aims to address this gap.

We consider a susceptible-infected-susceptible (SIS) epidemic model with two virus strains, denoted by $\Ht$ and $\Lt$, each with distinct transmission and recovery rates, where strain $\Ht$ exhibits higher reproduction number and transmission rate in comparison to strain $\Lt$. Individuals can only be infected by one strain at any given time. Our game setting models all susceptible individuals as players who choose between adopting protection and remaining unprotected to maximize their individual payoffs. 

Our analysis begins with the general case of bi-directional mutation, where individuals infected by one strain can mutate to the other. The mutation from $\Lt$ to $\Ht$ may arise from selective advantages, whereas, mutation from $\Ht$ to $\Lt$ may result from deleterious mutations such as replication errors. We show that the epidemic dynamics with mutations and protection adoption always admits a disease-free equilibrium which is either globally asymptotically stable in a suitable parameter regime or unstable otherwise. We then establish the existence of a unique endemic equilibrium of coexisting strains and prove its global asymptotic stability under suitable assumptions. Furthermore, we characterize the stationary Nash equilibrium (SNE) of the game-theoretic model, and show that infection levels decrease monotonically with increased protection adoption.

We then investigate the case of uni-directional mutation, considering two scenarios: mutation \emph{only} from $\Lt$ to $\Ht$ and mutation \emph{only} from $\Ht$ to $\Lt$. Our results reveal that in the first scenario, coexistence of both strains is impossible, reducing the system to a single-virus setting. In contrast, in the second scenario, both strains coexist at steady-state. We characterize the SNEs when both viruses coexist. We analytically show that when the mutation rate is below a threshold, both strains persist; however, as the mutation rate increases, the prevalence of $\Ht$ decreases, and beyond the threshold, a single-strain endemic equilibrium dominated by $\Lt$ emerges. We validate our major findings through numerical simulations.

The rest of the paper is organized as follows. Section \ref{sec:mut_strain_l_to_h} introduces the SIS epidemic dynamics with protection adoption and mutation. Section \ref{sec:bidir_mut} presents a detailed analysis of equilibria under bi-directional mutation. Characterization of the SNE under bi-directional mutation is presented in Section \ref{sec:char:sne:bidir}. Uni-directional mutation from the dominant strain is examined in Section \ref{sec:steadystate_muta_strain_h_to_l}, followed by the SNE characterization. Section \ref{sec:mut_to_dom_strain} gives a brief discussion on mutation to the dominant strain. Numerical results are included in Section \ref{sec:num_sim}. We conclude the paper in Section \ref{sec:conc}, along with possible directions of future work. 

%In the remaining part of the section, we introduce some notations.

\subsection{Notation and Background}
\label{subsec:notation}

The stability modulus of an $n$-dimensional matrix $A$ denoted by $\phi(A)$, and the spectral radius denoted by $\rho(A)$ are defined as:
\begin{align*}
    \phi(A) &:= \max_{1 \leq i \leq n} \{\; \mathtt{Real} (\lambda_{i}) \; \}, 
    \\ \rho(A) &:= \max_{1 \leq i \leq n} \{\; \mid \lambda_{i} \mid \; \},
\end{align*}
where $\lambda_{i}$ is the $i$-th eigenvalue of the matrix, and $\mathtt{Real}(\cdot)$ returns the real component of its argument. We introduce two real-valued functions, each of which inputs a matrix as its argument. $\mathtt{Tr}\big[\cdot \big]$ computes the trace of a matrix, whereas $\mathtt{Det}\big[\cdot \big]$ returns its determinant. In addition, $\Delta_{3} \subset \mathbb{R}^{3}$ denotes the probability simplex in three-dimension. Finally, we state the following proposition from \cite{liu2019analysis} which is used in some of our proofs.

\begin{proposition}[Proposition 1, \cite{liu2019analysis}]
    Suppose $\Lambda \in \mathbb{R}^{n \times n}$ is a diagonal matrix with all entries being strictly negative, and $N \in \mathbb{R}^{n \times n}$ is an irreducible nonnegative matrix. Let $M = \Lambda + N$. Then, following relations hold:
    \begin{itemize}
        \item $\phi(M) < 0 \iff \rho (-{\Lambda}^{-1} N) < 1$,
        \item $\phi(M) = 0 \iff \rho (-{\Lambda}^{-1} N) = 1$, 
        \item $\phi(M) > 0 \iff \rho (-{\Lambda}^{-1} N) > 1$.
    \end{itemize}
    \label{prop:pro1_liu_pare}
\end{proposition}

}

\section{SIS Mutation Model Coupled with Protection Adoption Behavior}
\label{sec:mut_strain_l_to_h}

We examine the Susceptible-Infected-Susceptible (SIS) epidemic model, where individuals can be in one of three states: susceptible, infected with strain $\Ht$, and infected with strain $\Lt$. The transmission and recovery rates of strain $\Ht$ are denoted by $\betah, \gammah > 0$, while the rates for strain $\Lt$ are denoted by $\betal, \gammal > 0$. We assume that $\frac{\betah}{\gammah} > \frac{\betal}{\gammal}$ and $\betah > \betal$, i.e., $\Ht$ has a stronger infection rate and reproduction number compared to $\Lt$. We denote the fraction of the population that is susceptible, infected by strain $\Ht$, and infected by strain $\Lt$ by $\St, \Ih$ and $\Il$, respectively. An individual can be infected by only one strain at a time, i.e., we exclude the possibility of simultaneous infection by both strains. 

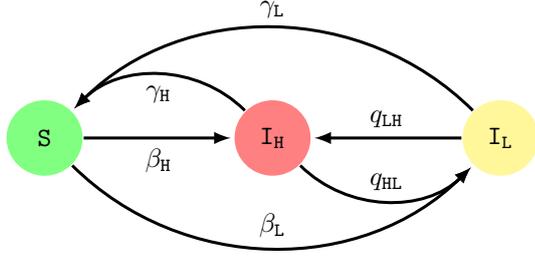
\begin{figure}[t!]
\centering
\begin{tikzpicture}
% Setup the style for the states
\tikzset{node style/.style={circle, minimum width=1cm, line width=0.3mm, fill=green!50!white}}
\tikzset{node style1/.style={circle, minimum width=1cm, line width=0.3mm, fill=red!50!white}}
\tikzset{node style2/.style={circle, minimum width=1cm, line width=0.3mm, fill=yellow!50!white}}
        % Draw the states
\node[node style] at (0, 0) (St)     {$\St$};
\node[node style1] at (3, 0) (Ih)     {$\Ih$};
\node[node style2] at (6, 0) (Il)  {$\Il$};
        % Connect the states with arrows
\draw[every loop, auto=right,line width=0.4mm,
              >=latex,
              draw=black,
              fill=black]
            %(St)     edge[bend right=20]            node {0.1} (Yt)
(St) edge[bend right=0, auto=left] node[below] {$\betah$} (Ih)
(St) edge[bend right=45, auto=left] node[above] {$\betal$} (Il)
(Ih) edge[bend right=45] node[below] {$\gammah$} (St)
(Il) edge[bend right=0] node[above] {$\qlh$} (Ih)
(Il) edge[bend right=45] node[above] {$\gammal$} (St)
(Ih) edge[bend right=45] node[above] {$\qhl$} (Il);
\end{tikzpicture}
\caption{Evolution of infection states of an individual in the SIS model with two mutant strains.}
\label{fig:sis_mut_l_to_h}
\end{figure}

We focus on the mutation process that drives individuals infected with one strain to transition to the other strain. We introduce mutation rate $\qhl \geq 0$ to denote the transition rate from strain $\Ht$ to strain $\Lt$, whereas $\qlh \geq 0$ captures the mutation rate from strain $\Lt$ to strain $\Ht$. The mutation from strain $\Lt$ to strain $\Ht$ accounts for the transition to the dominating strain (similar to \cite{bonhoeffer1994mutation}), whereas mutation from $\Ht$ to $\Lt$ is potentially due to replication error \cite{sanjuan2016mechanisms}. The transition between different disease states is represented in Figure \ref{fig:sis_mut_l_to_h}. Note that our setting differs from bi-virus models studied in most of the past works, such as \cite{pare2021multi,zhang2022networked,watkins2015optimal}, which do not allow direct transition from one infection state to another. 

We now incorporate protection adoption behavior of individuals in the SIS epidemic dynamics with virus mutation. We assume that each susceptible agent chooses among two available actions: adopting protection and remaining unprotected. Formally, we denote $a \in \{\Pt, \Ut\}$. For an unprotected susceptible agent, the rate of infection by strain $\Ht$ (respectively, strain $\Lt$) is $\betah \Ih$ (respectively, $\betal \Il$). Adopting protection reduces the likelihood of becoming infected by both the strains by a factor $\alpha \in (0, 1)$. The proportion of susceptible individuals that adopt protection is denoted by $\zs \in [0,1]$; this quantity depends on the payoff functions that are introduced subsequently. 

The SIS epidemic dynamics with mutation and game-theoretic protection adoption are given by
 \begin{align}
    &\dot{\St}(t) = -(\betah \Ih(t) + \betal \Il(t)) \big(\alpha \zs(t) + 1 - \zs(t)\big) \St(t) \nonumber
    \\ & \qquad \qquad + \gammah \Ih(t) + \gammal \Il(t), \nonumber
    \\ &\dot{\mathtt{I}}_{\mathtt{H}}(t) = \betah \Ih(t) \big(\alpha \zs(t) + 1 - \zs(t)\big) \St(t) - \qhl \Ih(t) \nonumber
    \\ & \qquad \qquad + \qlh \Il(t) - \gammah \Ih(t), \nonumber
    \\ &\dot{\mathtt{I}}_{\mathtt{L}}(t) = \betal \Il(t) \big(\alpha \zs(t) + 1 - \zs(t)\big) \St(t) + \qhl \Ih(t) \nonumber
    \\ & \qquad \qquad -\qlh \Il(t) - \gammal \Il(t).
\label{eq:sys_dyn}
\end{align}

From dynamics \eqref{eq:sys_dyn}, the relation $\dot{\St}(t) + \dot{\mathtt{I}}_{\mathtt{H}}(t) + \dot{\mathtt{I}}_{\mathtt{L}}(t) = 0$ holds for all $t$. We first focus on analyzing the equilibria of \eqref{eq:sys_dyn} in the following section. 

\section{Equilibria and their Stability under Bi-directional Mutation}
\label{sec:bidir_mut}

We first establish the invariance of $\Delta_3$ for the above dynamics in the following lemma. 

\begin{lemma}
    Given $\zs \in [0, 1]$, the set
    \begin{equation*}
        \Delta_3 := \bigg\{(\St, \Ih, \Il) \in [0, 1]^3 \bigg|\St+ \Ih+ \Il=1 \bigg\},  
    \end{equation*}
    is positive invariant with respect to \eqref{eq:sys_dyn}.
    \label{lemm:positive_invariant_bidir}
\end{lemma}

\begin{proof}
    We assume that $\St(0), \Ih(0), \Il(0) \in \Delta_3$. Since, $\dot{\St}(t) + \dot{\mathtt{I}}_{\mathtt{H}}(t)  + \dot{\mathtt{I}}_{\mathtt{L}}(t) = 0$ is true for all $t \geq 0$, we conclude that $\St(t) + \Ih(t) + \Il(t) = 1$ is satisfied for all $t > 0$. It can be shown that every initial condition in the space of $\Delta_{3}$ to the ODE defined by \eqref{eq:sys_dyn} admits a unique solution (see \cite[Chapter 2]{bressan2007introduction}). 
    
    We now analyze the states governed by \eqref{eq:sys_dyn} at boundary of the simplex $\Delta_{3}$. At any time $t \geq 0$, if $\Ih(t) = 0$, then $\dot{\mathtt{I}}_{\mathtt{H}}(t) = \qlh \Il \geq 0$. When $\Ih(t) = 1$, we obtain $\dot{\mathtt{I}}_{\mathtt{H}}(t) < 0$, since $\qhl, \gammah > 0$ and $\St(t)=0$. This proves that proportion infected by strain $\Ht$ is confined to $\Ih(t) \in [0, 1]$. 
    
    Similarly, when $\Il(t) = 0$, we obtain $\dot{\mathtt{I}}_{\mathtt{L}}(t) = \qhl \Ih \geq 0$, and when $\Il(t) = 1$, its derivative $\dot{\mathtt{I}}_{\mathtt{L}}(t) = -(\qlh + \gammal) \Il < 0$ since $\St(t)=\Ih(t)=0$. The result now follows from Nagumo's theorem \cite{blanchini2008set}.
\end{proof}

With the positive invariance of the simplex being established, we proceed with the analysis of the disease-free equilibrium.
\subsection{Stability of Disease-free Equilibrium}

From the above discussion, we have $\dot{\St} = -\dot{\mathtt{I}}_{\mathtt{H}} - \dot{\mathtt{I}}_{\mathtt{L}}$ and $\St = 1 - \mathtt{I}_\mathtt{H} - \mathtt{I}_{\mathtt{L}}$. Hence, we can rewrite the dynamics in \eqref{eq:sys_dyn} as:
\begin{subequations}
\begin{align}
     &\dot{\mathtt{I}}_{\mathtt{H}}(t) = \hatbetah(\zs(t)) \Ih(t) (1 - \Ih(t) - \Il(t)) + \qlh \Il(t) \nonumber
     \\ &\qquad \qquad - (\qhl + \gammah) \Ih(t), 
     \label{eq:bidir_ihdot}
    \\ &\dot{\mathtt{I}}_{\mathtt{L}}(t) = \hatbetal(\zs(t)) \Il(t) (1 - \Ih(t) - \Il(t)) + \qhl \Ih(t) \nonumber
    \\ &\qquad \qquad - (\qlh + \gammal) \Il(t),
\label{eq:bidir_ildot}
\end{align}
\end{subequations}
where $\hat{\beta}_{\mathtt{Q}}(\zs(t)) := \beta _{\mathtt{Q}} (\alpha \zs(t) + 1 - \zs(t))$, for $\mathtt{Q} \in \{\mathtt{H}, \mathtt{L}\}$. Note from \eqref{eq:bidir_ihdot} and \eqref{eq:bidir_ildot} that the disease-free equilibrium always exists. We denote the disease-free equilibrium by $\textbf{E}_{\mathtt{DFE}} := (\Ststar = 1, \Ihstar = 0, \Ilstar = 0)$, and analyze its stability. 

We rewrite \eqref{eq:bidir_ihdot} and \eqref{eq:bidir_ildot} in matrix form as:
\begin{equation}
    \dot{\textbf{x}} = (B - D + M) \textbf{x} - X B \;\mathbb{1}\; \textbf{x},
    \label{eq:matrix_dynamics}
\end{equation}
where, $\textbf{x} = \begin{bmatrix}
    \Ih\\
    \Il
\end{bmatrix} \in [0, 1]^{2}$ is the state vector, $X \in \mathbb{R}^{2 \times 2}$ is the diagonal matrix with $\Ih$ and $\Il$ as the diagonal elements, $\mathbb{1} \in \mathbb{R}^{2 \times 2}$ has all its elements as one, and matrices $B = \begin{bmatrix}
            \hatbetah & 0
            \\ 0 & \hatbetal
        \end{bmatrix}$, $D = \begin{bmatrix}
            \gammah + \qhl & 0
            \\ 0 & \gammal + \qlh
        \end{bmatrix}$, $M = \begin{bmatrix}
            0 & \qlh
            \\ \qhl & 0
         \end{bmatrix}$. We now state the following theorem.
\begin{theorem}
    The disease-free equilibrium, $\textbf{E}_{\mathtt{DFE}} := (1, 0, 0)$ is globally asymptotically stable if and only if $\rho\big(D^{-1} (B + M) \big) \leq 1$.
    \label{thm:stab_dfe}
\end{theorem}
\begin{proof}
    We begin the proof by computing the Jacobian at DFE as
\begin{align}
            \mathcal{J}(\textbf{E}_{\mathtt{DFE}}) &:= 
            \begin{bmatrix}
            \hatbetah - \gammah - \qhl & \qlh
            \\ \qhl & \hatbetal - \gammal - \qlh
        \end{bmatrix} \label{eq:jac_dfe}
        \\&= B - D + M. \nonumber
        \end{align}
 Observe that $-D$ is a negative diagonal matrix, whereas $B + M$ is an irreducible nonnegative matrix. Following Proposition \ref{prop:pro1_liu_pare}, the necessary and sufficient condition for the DFE to be locally stable is that the stability modulus satisfies $\phi(B - D + M) \leq 0$, which in turn implies that the spectral radius is $\rho\big(D^{-1} (B + M) \big) \leq 1$. We denote $\mathcal{R}_{0} := \rho\big(D^{-1} (B + M) \big)$ as the reproduction number, consistent with the literature of epidemics.

Note that when the spectral radius $\rho\big(D^{-1} (B + M) \big) > 1$, from the computed Jacobian $\mathcal{J}(\textbf{E}_{\mathtt{DFE}})$ it can be shown that the disease-free equilibrium is unstable. Thus, $\rho\big(D^{-1} (B + M) \big) \leq 1$ is necessary to ensure the stability of the equilibrium. 

We now derive the sufficient condition for stability by exploiting a special property of Metzler matrices. Since, the Jacobian matrix $\mathcal{J}(\textbf{E}_{\mathtt{DFE}}) = B - D + M$ has all non-negative off-diagonal elements, it is a Metzler matrix, and there exists a positive diagonal matrix $P^{\mathtt{DFE}}$ such that
\begin{equation*}
    (B - D + M)^{\top} P^{\mathtt{DFE}} + P^{\mathtt{DFE}} (B - D + M) = -K^{\mathtt{DFE}},
\end{equation*}
where $K^{\mathtt{DFE}}$ is a positive semi-definite matrix \cite[Lemma A.1]{khanafer2016stability}, \cite{rantzer2011distributed}. We consider $V(\textbf{x}) = \textbf{x}^{\top} P^{\mathtt{DFE}} \textbf{x}$ as a candidate Lyapunov function. The time-derivative of $V(\textbf{x})$ yields
\begin{align}
    \dot{V} &= \textbf{x}^{\top} P^{\mathtt{DFE}} \dot{\textbf{x}} + \dot{\textbf{x}}^{\top} P^{\mathtt{DFE}} \textbf{x} \nonumber
    \\ &= \textbf{x}^{\top} \big( P^{\mathtt{DFE}} (B - D + M) + (B - D + M)^{\top} P^{\mathtt{DFE}} \big) \textbf{x} \nonumber
    \\ & \qquad- \textbf{x}^{\top} P^{\mathtt{DFE}} X B \;\mathbb{1}\; \textbf{x} - \textbf{x}^{\top} \;\mathbb{1}\; P^{\mathtt{DFE}} X B \textbf{x} \nonumber
    \\ &= -\textbf{x}^{\top} K^{\mathtt{DFE}} \textbf{x} - \textbf{x}^{\top} \underbrace{P^{\mathtt{DFE}} X B}_{\succeq 0} \;\mathbb{1}\; \textbf{x} \nonumber
    \\&\qquad - \textbf{x}^{\top} \;\mathbb{1}\; P^{\mathtt{DFE}} X B \textbf{x} \leq 0. \label{eq:Vdot_dfe}
\end{align}
Note that $P^{\mathtt{DFE}} X B = B X P^{\mathtt{DFE}}$ holds, since $P^{\mathtt{DFE}}, X$ and $B$ all are diagonal matrices with non-negative entries. Finally we prove that $\dot{V} = 0$ if and only if $\textbf{x} = \textbf{0}$ holds. A simple expansion of the second and third terms in \eqref{eq:Vdot_dfe} results in 
\begin{align*}
     &-\textbf{x}^{\top} P^{\mathtt{DFE}} X B \;\mathbb{1}\; \textbf{x} - \textbf{x}^{\top} \;\mathbb{1}\; P^{\mathtt{DFE}} X B \textbf{x} 
    \\ &=- \Ih \big( 2 p_{1}^{\mathtt{DFE}} \Ih^2 \hatbetah + \Il (p_{1}^{\mathtt{DFE}} \Ih \hatbetah + p_{2}^{\mathtt{DFE}} \Il \hatbetal) \big) 
    \\ &\quad - \Il \big( 2 p_{2}^{\mathtt{DFE}} \Il^2 \hatbetal + \Ih (p_{1}^{\mathtt{DFE}} \Ih \hatbetah + p_{2}^{\mathtt{DFE}} \Il \hatbetal) \big),
\end{align*}
where $p_{1}^{\mathtt{DFE}}$ and $p_{2}^{\mathtt{DFE}}$ are the positive diagonal elements of matrix $P^{\mathtt{DFE}}$. Note that the above expression is strictly negative for $\big(\Ih \in (0, 1], \Il \in [0, 1] \big) \text{ and }\big(\Ih \in [0, 1], \Il \in (0, 1] \big)$, and equals zero if and only if $\Ih = \Il = 0$ is true. Since $K^{\mathtt{DFE}}$ in \eqref{eq:Vdot_dfe} is a positive semi-definite matrix, it is clear that $\dot{V} < 0$ holds only when either of the infection states is strictly positive, and $\dot{V} = 0$ holds if and only if $\Ih = \Il = 0$ is satisfied. Thus, the disease-free equilibrium $\textbf{E}_{\mathtt{DFE}}$ is globally asymptotically stable. This concludes our proof.
\end{proof}

In the following section, we analyze the existence and stability of an endemic equilibrium of \eqref{eq:sys_dyn}.  

\subsection{Characterization of Endemic Equilibrium}

For a given $\zs \in [0, 1]$, we denote an endemic equilibrium of \eqref{eq:bidir_ihdot} and \eqref{eq:bidir_ildot} by $\textbf{E}_{\mathtt{EE}}(\zs) := (\Ihstar(\zs), \Ilstar(\zs))$ with both $\Ihstar(\zs), \Ilstar(\zs) \in (0,1)$. Throughout the section, we suppress the dependence on $\zs$ for brevity, and explicitly include it when necessary. We begin this section by first showing that if an endemic equilibrium exists, then it is locally stable.
\begin{lemma}
    Let $\textbf{E}_{\mathtt{EE}} = (\Ihstar, \Ilstar)$ with $\Ihstar, \Ilstar \in (0, 1)^2$ be an endemic equilibrium of \eqref{eq:bidir_ihdot} and \eqref{eq:bidir_ildot}. Then $\textbf{E}_{\mathtt{EE}}$ is locally stable.
    \label{lemm:stab_ee}
\end{lemma}
\begin{proof}
    By rearrangement of the terms in \eqref{eq:bidir_ihdot} and \eqref{eq:bidir_ildot}, we obtain the following equations:
\begin{align}
    \hatbetah \Ststar -\gammah - \qhl = -\frac{\qlh \Ilstar}{\Ihstar}, \;\; \hatbetal \Ststar -\gammal - \qlh = -\frac{\qhl \Ihstar}{\Ilstar}.
    \label{eq:bidir_dot_ih_il=0}
\end{align}
%\vspace{-2mm}
The Jacobian at any endemic equilibrium $\textbf{E}_{\mathtt{EE}}$ of the dynamics \eqref{eq:bidir_ihdot} and \eqref{eq:bidir_ildot} is given by
\begin{align}
            &\mathcal{J}(\textbf{E}_{\mathtt{EE}}) := \nonumber
            \\ &\begin{bmatrix}
            \hatbetah (1 \! - \!2 \Ihstar\! -\! \Ilstar) \!-\! \gammah \!-\! \qhl & -\!\hatbetah \Ihstar + \qlh
            \\ -\!\hatbetal \Ilstar \!+\! \qhl & \hatbetal (1\! -\! \Ihstar\! -\! 2 \Ilstar) \!-\! \gammal \! - \! \qlh
        \end{bmatrix} \nonumber
        \\=&\begin{bmatrix}
            -\frac{\qlh \Ilstar}{\Ihstar} - \hatbetah \Ihstar & -\hatbetah \Ihstar + \qlh
            \\ -\hatbetal \Ilstar + \qhl & -\frac{\qhl \Ihstar}{\Ilstar} - \hatbetal \Ilstar
        \end{bmatrix},
        \label{eq:jac_ee}
\end{align}
where \eqref{eq:jac_ee} is obtained by leveraging \eqref{eq:bidir_dot_ih_il=0}. From \eqref{eq:jac_ee}, we have
\begin{align*}
    \mathtt{Tr}&\big[\mathcal{J}(\textbf{E}_{\mathtt{EE}}) \big] = -\frac{\qlh \Ilstar}{\Ihstar} - \hatbetah \Ihstar -\frac{\qhl \Ihstar}{\Ilstar} - \hatbetal \Ilstar < 0,
    \\\mathtt{Det}&\big[\mathcal{J}(\textbf{E}_{\mathtt{EE}}) \big] = \bigg(\frac{\qlh \Ilstar}{\Ihstar} + \hatbetah \Ihstar\bigg) \bigg(\frac{\qhl \Ihstar}{\Ilstar} + \hatbetal \Ilstar \bigg) 
    \\ &\!\!\!\!-(-\hatbetah \Ihstar + \qlh) (-\hatbetal \Ilstar + \qhl),
    \\ &\!\!\!\! =\frac{\qlh \hatbetal (\Ilstar)^{2}}{\Ihstar} + \frac{\qhl \hatbetah (\Ihstar)^{2}}{\Ilstar} + \qhl \hatbetah \Ihstar + \qlh \hatbetal \Ilstar > 0.
\end{align*}
The above inequalities show that both eigenvalues of $\mathcal{J}(\textbf{E}_{\mathtt{EE}})$ are strictly negative, implying that an endemic equilibrium $\textbf{E}_{\mathtt{EE}}$ is locally stable, whenever it exists. 
\end{proof}

\begin{remark}
    Note that when the reproduction number satisfies $\mathcal{R}_{0} \leq 1$, by Theorem \ref{thm:stab_dfe} the disease-free equilibrium exhibits global stability. Again, Lemma \ref{lemm:stab_ee} states that existence of an endemic equilibrium implies that the equilibrium is also locally stable. Therefore, we conclude that an endemic equilibrium does not exist when $\mathcal{R}_{0} \leq 1$ holds.
    \label{remark:ee_not_exist}
\end{remark}

 We next prove the existence and uniqueness of an endemic equilibrium under the following assumption.
 \begin{assumption}
     For a fixed $\zs \in [0, 1]$, the mutation rates satisfy the following inequality: $$\max(\qlh + \gammal, \qhl + \gammah) \leq \min \big(\hatbetah(\zs), \hatbetal(\zs) \big).$$
     \label{assump:mutation_rates}
 \end{assumption}
 %\vspace{-8mm}
 \begin{remark}
     The above assumption implies that effective transmission rates of both strains are not smaller than the sum of the mutation and recovery rates of both strains. Indeed for most infectious diseases, infection rates are larger compared to the sum of the rates at which they mutate and recover.

    Observe that when Assumption \ref{assump:mutation_rates} holds, sum of the eigenvalues of Jacobian $\mathcal{J}(\textbf{E}_{\mathtt{DFE}})$ in \eqref{eq:jac_dfe} is positive, which implies that the largest eigenvalue is positive. Consequently, when the parameters satisfy Assumption \ref{assump:mutation_rates} at some $\zs \in [0,1]$, we have $\mathcal{R}_{0}(\zs) > 1$.
 \end{remark}    
 
 \begin{theorem}
     Suppose Assumption \ref{assump:mutation_rates} holds for a given $\zs \in [0, 1]$. Then the system defined by \eqref{eq:bidir_ihdot} and \eqref{eq:bidir_ildot} admits a unique endemic equilibrium. 
     \label{thm:existence}
 \end{theorem}
\begin{proof}
    The proof is presented in Appendix \ref{sec:appen_existence}.
\end{proof}

We now present the main theorem establishing the global asymptotic stability of the endemic equilibrium.

 \begin{theorem}
     Suppose Assumption \ref{assump:mutation_rates} holds for a given $\zs \in [0, 1]$. Then, the unique endemic equilibrium $\Ihstar(\zs), \Ilstar(\zs) \in (0,1)$ is (almost) globally asymptotically stable. 
     \label{thm:stab_ee}
 \end{theorem}
 \begin{proof}
     Recall that when Assumption \ref{assump:mutation_rates} is satisfied, $\mathcal{R}_{0}(\zs) > 1$ holds. Then,  disease-free equilibrium is unstable, and a unique locally stable endemic equilibrium exists. We begin the proof of global asymptotic stability of the endemic equilibrium by ruling out the existence of any closed orbits. We omit the dependence on $\zs$ is the remainder of the proof.
     
     Our dynamics has two independent states, the dynamics of both are continuously differentiable on the domain of $(0, 1)^2$. We define the real-valued function $h: (0, 1)^2 \rightarrow \mathbb{R}$, such that, $h(\textbf{x}) := \frac{1}{\Ih \Il}$ where $\textbf{x} = (\Ih, \Il)$. Note that domain of the mapping $h(\textbf{x})$ excludes the disease-free equilibrium, and it is continuously differentiable on $(0, 1)^2$. Calculating the product $$h(\textbf{x}) \; \dot{\textbf{x}} = \begin{bmatrix}
         \frac{\hatbetah}{\Il} (1 - \Ih - \Il) + \frac{\qlh}{\Ih} - \frac{(\qhl + \gammah)}{\Il} \\
        \frac{\hatbetal}{\Ih} (1 - \Ih - \Il) + \frac{\qhl}{\Il} - \frac{(\qlh + \gammal)}{\Ih}
     \end{bmatrix},$$ and computing its divergence, we obtain $$\nabla \cdot (h \; \dot{\textbf{x}}) = -\frac{\hatbetah}{\Il} - \frac{\qlh}{\Ih^2} - \frac{\hatbetal}{\Ih} - \frac{\qhl}{\Il^2} < 0,$$ throughout the domain of $(0, 1)^2$, i.e., the sign of $\nabla \cdot (h \; \dot{\textbf{x}})$ remains unchanged (i.e., remains negative). Therefore, by leveraging Dulac's Criterion \cite{strogatz2024nonlinear}, we conclude that closed orbits do not exist in the space of $(0, 1)^2$.

     Consequently, in the absence of any closed orbits, the disease-free equilibrium being unstable, and the unique endemic equilibrium being locally stable in the two-dimensional system defined by \eqref{eq:bidir_ihdot} and \eqref{eq:bidir_ildot}, the only possibility left for any trajectory with any initial conditions in the space of $(0, 1)^2$ is to converge to the endemic equilibrium and remain there (which exhibits local stability).
 \end{proof}

\section{Game-Theoretic Protection Adoption under Bi-directional Mutation}
\label{sec:char:sne:bidir}

We now integrate the mutation-driven epidemic propagation and the game-theoretic strategies of protection adoption. First, we introduce the rewards for susceptible individuals. As mentioned earlier, a susceptible individual has the choice of either adopting protection or remaining unprotected. It incurs a cost of $\cp > 0$ on adopting protection, whereas on remaining unprotected there is no cost involved. We define the instantaneous reward received by a susceptible agent choosing an action $a \in \{\Pt,\Ut\}$ as
 \begin{align}
     R[\Pt](\Ih,\Il) &= -\cp - \alpha \big(\rhoh \betah \Ih + \rhol \betal \Il \big), \nonumber
     \\ R[\Ut](\Ih,\Il) &= - \big(\rhoh \betah \Ih + \rhol \betal \Il \big), \label{eq:rewards}
 \end{align}
where $\rhoh$ (respectively, $\rhol$) captures the loss a susceptible agent incurs upon infection by strain $\Ht$ (respectively, strain $\Lt$). We further define
\begin{align}
    \Delta R(\Ih,\Il) &:= R[\Pt](\Ih,\Il) -R[\Ut](\Ih,\Il) \nonumber
    \\ & = -\cp + (1-\alpha) \big(\rhoh \betah \Ih + \rhol \betal \Il \big). \label{eq:reward_difference}
\end{align}
We now provide a formal definition of \emph{stationary Nash equilibrium} (SNE) of this game.
 
\begin{definition}\label{def:sne}
The tuple $(\Ihstar, \Ilstar, \zsne)$ is a Stationary Nash equilibrium if $(\Ihstar, \Ilstar)$ denotes the stable equilibrium point of \eqref{eq:bidir_ihdot} and \eqref{eq:bidir_ildot} at $\zsne$, and the proportion that adopts protection $\zsne$ satisfies the following conditions: 
\small{\begin{align*}
& \zsne = 0 \Rightarrow \Delta R(\Ihstar,\Ilstar) \leq 0, \text{ and } \Delta R(\Ihstar,\Ilstar) < 0 \Rightarrow \zsne = 0,
\\ & \zsne = 1 \Rightarrow \Delta R(\Ihstar,\Ilstar) \geq 0, \text{ and } \Delta R(\Ihstar,\Ilstar) > 0 \Rightarrow \zsne = 1,
\\ & \zsne \in (0,1) \Rightarrow \Delta R(\Ihstar,\Ilstar) = 0. %, \text{ and } 
%\\ & \qquad  \Delta R(\Ihstar,\Ilstar) = 0 \Rightarrow  \zsne \in [0,1].
\end{align*}}
\end{definition}
 
We first state the following lemma which is crucial for the characterization of SNE.

\begin{lemma}\label{prop:states_mon_dec}
The infected proportions at the endemic equilibrium, $\Ihstar(\zs)$ and $\Ilstar(\zs)$, are monotonically decreasing in the proportion that adopt protection $\zs$.
\end{lemma}
\begin{proof}
    The proof is included in Appendix \ref{sec:appen_states_mon_dec}.
\end{proof}

We now define
\begin{align}
    \cmin &:= (1 - \alpha) \big(\betah \rhoh \Ihstar(\zs = 1) + \betal \rhol \Ilstar(\zs = 1)\big), \nonumber
    \\ \cmax &:= (1 - \alpha) \big(\betah \rhoh \Ihstar(\zs = 0) + \betal \rhol \Ilstar(\zs = 0)\big).
    \label{eq:imin_imax}
\end{align}
Our main theorem pertaining to the characterization of SNE is stated below.

\begin{theorem}\label{theorem:char_NE_bidir}
Suppose Assumption \ref{assump:mutation_rates} holds for $\zs = 1$. The following statements hold for the protection behavior $\zsne$ at stationary Nash equilibria:
     \begin{enumerate}[(a)]
         \item if $\cp < \cmin \text{, we have } \zsne=1$;
         \item if $\cp > \cmax \text{, we have } \zsne=0$;
         \item if $\cmin \leq \cp \leq \cmax$, then there exists a unique $\zsne \in [0, 1]$ such that $\Delta R(\Ihstar(\zsne),\Ilstar(\zsne)) = 0$. 
      \end{enumerate}
The infected proportions at the SNE are at their respective unique endemic equilibria given by $\Ihstar(\zsne),\Ilstar(\zsne)$. 
\end{theorem}

\begin{proof}
     Recall from the proof of Theorem \ref{thm:stab_dfe} that $$\mathcal{R}_{0}(\zs) = \rho \Bigg(\begin{bmatrix}
        \frac{\hatbetah(\zs)}{\gammah + \qhl} & \qlh\\
        \qhl & \frac{\hatbetal(\zs)}{\gammal + \qlh}
    \end{bmatrix} \Bigg).$$ 
    Since the above matrix is positive and irreducible, its spectral radius is monotonically increasing in each element of the matrix; a consequence of the Perron-Frobenius Theorem \cite[Corollary 8.1.19]{horn2012matrix}. As a result, $\mathcal{R}_{0}(\zs)$ is monotonically increasing in $\frac{\hatbetah(\zs)}{\gammah + \qhl}$ and $\frac{\hatbetal(\zs)}{\gammal + \qlh}$. However, $\hatbetah(\zs)$ and $\hatbetal(\zs)$ are both individually decreasing in $\zs$, which implies that $\mathcal{R}_{0}(\zs)$ is monotonically decreasing in $\zs$. Consequently, $\mathcal{R}_{0}(1) > 1 \Rightarrow \mathcal{R}_{0}(\zs) > 1, \; \forall \; \zs \in [0, 1)$, which implies that an endemic equilibrium exists and is GAS for all $\zs$. 
    
    We now define the difference in rewards at the endemic equilibrium as 
     \begin{equation}
         \Delta R[\textbf{E}_{\mathtt{EE}}](\zs) := -\cp + (1 - \alpha) \big(\rhoh \betah \Ihstar (\zs) + \rhol \betal \Ilstar (\zs) \big).
         \label{eq:rew_diff_end_eq}
     \end{equation}
     We now analyze the Nash strategies at the endemic equilibrium. Recall from Lemma \ref{prop:states_mon_dec} that both $\Ihstar(\zs)$ and $\Ilstar(\zs)$ are monotonically decreasing in $\zs$. Thus, from \eqref{eq:rew_diff_end_eq} we see that $\Delta R[\textbf{E}_{\mathtt{EE}}](\zs)$ is monotonically decreasing in $\zs$, satisfying
     \begin{align}
         -\cp + \cmin \leq \Delta R[\textbf{E}_{\mathtt{EE}}](\zs) \leq -\cp + \cmax,
         \label{eq:delR_ee_bounds}
     \end{align}
     where $\cmin$ and $\cmax$ are defined in \eqref{eq:imin_imax}. We now derive the conditions under the three sub-cases. 
     
     \noindent \underline{\textbf{Case (a):} $\zsne = 1$}
     \vspace{1mm}
     \\When $\cp < \cmin$ then we obtain $\Delta R[\textbf{E}_{\mathtt{EE}}](\zs) > 0$ for all $\zs \in [0,1]$. Therefore, every individual strictly prefers to adopt protection irrespective of the strategies chosen by others.  Thus, $\zsne = 1$ is the only strategy that arises at the SNE. 

     For necessity, let $\zsne = 1$. Then, we must have
     $$\Delta R[\textbf{E}_{\mathtt{EE}}](1) = -\cp + \cmin \geq 0,$$
     since no individual would prefer to remain unprotected when everyone else adopt protection. 
     
     \noindent \underline{\textbf{Case (b):} $\zsne = 0$}
     \vspace{1mm}
     \\When $\cp > \cmax$ then we obtain $\Delta R[\textbf{E}_{\mathtt{EE}}](\zs) < 0$, which implies that $\zsne = 0$ is the only strategy that arises at the SNE following the reasoning stated above. 

     For necessity, let $\zsne = 0$. Then, we must have
     $$\Delta R[\textbf{E}_{\mathtt{EE}}](0) = -\cp + \cmax \leq 0,$$
     since no individual would prefer to adopt protection when no one else is doing so.

     \noindent \underline{\textbf{Case (c):} $\zsne \in [0, 1]$}
     \vspace{1mm}

    When the protection cost satisfies $\cmin \leq \cp \leq \cmax$, we observe that 
    $$\Delta R[\textbf{E}_{\mathtt{EE}}](1) \leq 0 \text{, and } \Delta R[\textbf{E}_{\mathtt{EE}}](0) \geq 0.$$ 
    Since $\Delta R[\textbf{E}_{\mathtt{EE}}](\zs)$ in $\zs$ is monotonically decreasing in $\zs$, there exists a unique $\zsne \in [0, 1]$ such that $\Delta R[\textbf{E}_{\mathtt{EE}}](\zsne) = 0$ holds which constitutes the SNE. In particular, an individual is indifferent between adopting protection or remaining unprotected, and hence, does not derive a strictly larger utility upon unilaterally changing its action.  
    
    This concludes the proof. 
  \end{proof}   
  
\begin{remark}
    The above theorem characterizes the protection behavior adopted by susceptible agents. When the protection cost exceeds the upper threshold, all susceptible agents remain unprotected, and when the protection cost is below the lower threshold, all agents are incentivized to adopt protection. For an intermediate protection cost, a unique non-zero fraction of the susceptible population adopts protection, while the rest remain unprotected.
    \label{rem:sne_char_bidir_remark}
\end{remark}

\begin{remark}
    We now briefly discuss the strategy adopted by susceptible agents when Assumption \ref{assump:mutation_rates} is not true for $\zs = 1$. From the monotonicity property, we know that if $\mathcal{R}_{0}(0) \leq 1 \Rightarrow \mathcal{R}_{0}(1) \leq 1$, and the endemic equilibrium does not exist. At the disease-free equilibrium, both strains are absent, i.e., $\Ihstar = \Ilstar = 0$ holds. Following \eqref{eq:reward_difference}, we have $\Delta R[\Ihstar,\Ilstar] = -\cp < 0$ when $\Ihstar = \Ilstar = 0$, i.e., for each individual, adopting protection has a smaller reward compared to being unprotected, irrespective of the strategies adopted by others. Thus, $\zsne = 0$ is the only Nash strategy, i.e., at the disease-free equilibrium, there exists a unique Nash strategy of remaining unprotected, which is intuitive as none of the viral strains survive.

    The setting where $\mathcal{R}_{0}(0) \geq 1$ and $\mathcal{R}_{0}(1) \leq 1$ could be analyzed along similar lines as the above theorem. However, it would only lead to more number of case analysis without significantly enhancing the technical contributions, and hence omitted. 
\end{remark}

With the complete characterization of the SNE in the general case of bi-directional mutation, we now consider the special case of mutation in a single direction, i.e., the mutation is either from strain $\Ht$ to strain $\Lt$, or from strain $\Lt$ to strain $\Ht$, but not in both directions.

\section{Specialization to Uni-directional Mutation}

We begin our analysis by allowing mutation only from strain $\Ht$ to $\Lt$.

\subsection{Mutation from Strain $\Ht$ to $\Lt$}
\label{sec:steadystate_muta_strain_h_to_l}

The dynamics of uni-directional mutation from strain $\Ht$ to $\Lt$ is obtained by setting $\qlh = 0$ and $\qhl > 0$ in \eqref{eq:bidir_ihdot} and \eqref{eq:bidir_ildot}, which yields
\um{\begin{align}
     &\dot{\mathtt{I}}_{\mathtt{H}}(t) = \hatbetah(\zs(t)) \Ih(t) \big(1 - \Ih(t) - \Il(t)\big) \nonumber
     \\ &\qquad \quad - \qhl \Ih(t) - \gammah \Ih(t), \nonumber
    \\ &\dot{\mathtt{I}}_{\mathtt{L}}(t) = \hatbetal(\zs(t)) \Il(t) \big(1 - \Ih(t) - \Il(t)\big) \nonumber
    \\ &\qquad \quad + \qhl \Ih(t)  - \gammal \Il(t).
\label{eq:mut_h_to_l}
\end{align}}

We now find out the equilibria of \eqref{eq:mut_h_to_l}, denoted by $(\St^\star,\Ih^\star,\Il^\star)$. To this end, we introduce the following notations:
    \begin{align}
        &w(\zs) := \alpha \zs + 1 - \zs, \nonumber
        \\ &\mathcal{D}_{1} := \betah \gammal - \betal \gammah + \betah \qhl - \betal \qhl, \nonumber
        \\ &\mathcal{N}^{\Ht}_{1}(\zs) := \gammah + \qhl - \hatbetah(\zs), \nonumber
        \\ &\mathcal{N}^{\Lt}_{1}(\zs) := \hatbetal(\zs) - \gammal.
        \label{eq:notations_n_d}
    \end{align}

The dynamics given by \eqref{eq:mut_h_to_l} has three equilibrium points for any given $\zs \in [0, 1]$, \um{the closed-form expressions of those are given below:}
\begin{align}
    \textbf{E}^{\Ht,\Lt}_{1} &= (1, 0, 0); \nonumber
     \\ \textbf{E}^{\Ht,\Lt}_{2}(\zs) &= \bigg(\frac{\gammal}{\hatbetal(\zs)}, 0, \frac{\mathcal{N}^{\Lt}_{1}(\zs)}{\hatbetal(\zs)}\bigg); \nonumber
     \\ \textbf{E}^{\Ht,\Lt}_{3}(\zs) &= \bigg(\frac{\gammah + \qhl}{\hatbetah(\zs)},  \label{eq:e_3}
     \\ & \qquad \frac{\mathcal{N}^{\Ht}_{1}(\zs) (-\mathcal{D}_{1} + \betah \qhl)}{\hatbetah(\zs) \mathcal{D}_{1}}, \frac{-\qhl \mathcal{N}^{\Ht}_{1}(\zs)}{w(\zs) \mathcal{D}_{1}} \bigg). \nonumber
\end{align}
Equilibrium $\textbf{E}^{\Ht,\Lt}_{1}$ is the disease-free equilibrium, whereas in $\textbf{E}^{\Ht,\Lt}_{2}(\zs)$ only strain $\Lt$ survives. Equilibrium $\textbf{E}^{\Ht,\Lt}_{3}(\zs)$ is the one in which both the strains co-exist. Observe that equilibria $\textbf{E}^{\Ht,\Lt}_{1}, \textbf{E}^{\Ht,\Lt}_{2}(\zs)$ and $\textbf{E}^{\Ht,\Lt}_{3}(\zs)$ are obtained as a limiting case of \eqref{eq:bidir_ihdot} and \eqref{eq:bidir_ildot} with $\qlh = 0$, such that, only one of the above equilibrium is stable under a given set of parameters as stated in the following corollary.

\begin{corollary}
    For a fixed $\zs \in [0, 1]$, the following statements hold for the GAS of equilibrium points.
    \begin{itemize}
        \item The disease-free equilibrium $\textbf{E}^{\Ht,\Lt}_{1}$ is GAS if and only if $\max\bigg(\frac{\hatbetah(\zs)}{\gammah + \qhl}, \frac{\hatbetal(\zs)}{\gammal} \bigg) < 1$;
        \item The single-strain equilibrium $\textbf{E}^{\Ht,\Lt}_{2}(\zs)$ is GAS if and only if the individual reproduction number of strain $\Lt$ is $\frac{\hatbetal(\zs)}{\gammal} > 1$, and the mutation rate satisfies $\qhl > \frac{\betah \gammal}{\betal} - \gammah$;
        \item The equilibrium of coexistence $\textbf{E}^{\Ht,\Lt}_{3}(\zs)$ is GAS if and only if $\frac{\hatbetah(\zs)}{\gammah + \qhl} > 1$ and the mutation rate satisfies $\qhl < \frac{\betah \gammal}{\betal} - \gammah$.
    \end{itemize}
    \label{cor:GAS_qlh=0}
\end{corollary}
\begin{proof}
    The proof is presented in Appendix \ref{sec:appen_corollary}. 
\end{proof}

A summary of the above results is included in Table \ref{table:summary}. 

\begin{table}[t!]
\renewcommand{\arraystretch}{2.5}
\setlength{\tabcolsep}{6pt}
\caption{Existence and stability of the equilibria of \eqref{eq:mut_h_to_l} for a given proportion of protection adoption $\zs \in [0,1]$ among susceptible individuals. Existence of an equilibrium is indicated by $\checkmark$, and if the equilibrium is stable, it is indicated with $\star$, whereas non-existence is denoted by $-$.}
\label{table:summary}
\centering
\begin{tabular}{|c | c | c | c |}
 \hline
 Parameters & $\textbf{E}^{\Ht,\Lt}_{1}(\zs)$ & $\textbf{E}^{\Ht,\Lt}_{2}(\zs)$ & $\textbf{E}^{\Ht,\Lt}_{3}(\zs)$\\ [0.5ex] 
 \hline \hline
 \parbox{3cm}{$\hatbetal(\zs) < \gammal$, and\\ $\hatbetah(\zs)<\gammah + \qhl$} & \checkmark, $\star$ & - & - \\ \hline
 \parbox{3cm}{$\hatbetal(\zs) > \gammal$,\\ $\hatbetah(\zs) < \gammah + \qhl$,\\and $\qhl > \frac{\betah \gammal}{\betal} - \gammah$} & \checkmark & \checkmark, $\star$ & - \\ \hline
 \parbox{3cm}{$\hatbetal(\zs) < \gammal$,\\ $\hatbetah(\zs) > \gammah + \qhl$,\\and $\qhl < \frac{\betah \gammal}{\betal} - \gammah$} & \checkmark & - & \checkmark, $\star$ \\ \hline
 \parbox{3cm}{$\hatbetal(\zs) > \gammal$, \\ $\hatbetah(\zs) > \gammah + \qhl$,\\and $\qhl > \frac{\betah \gammal}{\betal} - \gammah$} & \checkmark & \checkmark, $\star$ & - \\ \hline
 \parbox{3cm}{$\hatbetal(\zs) > \gammal$, \\ $\hatbetah(\zs) > \gammah + \qhl$,\\and $\qhl < \frac{\betah \gammal}{\betal} - \gammah$} & \checkmark  & \checkmark & \checkmark, $\star$ \\ \hline
\end{tabular}
\end{table}

\begin{remark}
The conditions $\frac{\hatbetal(\zs)}{\gammal} > 1$ and $\frac{\hatbetah(\zs)}{\gammah + \qhl} > 1$, depend on $\zs$ which varies in $[0, 1]$, whereas the inequality $\qhl > \frac{\betah \gammal}{\betal} - \gammah$ is independent of $\zs$. Thus, a low mutation rate ensures the survival of strain $\Ht$ with the equilibrium $\textbf{E}^{\Ht,\Lt}_{3}(\zs)$ being stable. As $\qhl$ increases, the proportion infected by strain $\Ht$ starts decreasing. When $\qhl$ is sufficiently large, we find a greater proportion of $\Ihstar$ transiting to $\Ilstar$ compared to the transition from $\Ststar$ to $\Ihstar$ which leads to strain $\Ht$ completely vanishing, which is observed in the regime that characterizes the stability of $\textbf{E}^{\Ht,\Lt}_{2}(\zs)$.
\end{remark}

Note that similar to Section \ref{sec:char:sne:bidir}, characterization of the SNE is also applicable when the virus mutates from strain $\Ht$ to $\Lt$. Substitution of the infection levels of $\Ihstar$ and $\Ilstar$ obtained from \eqref{eq:e_3} into \eqref{eq:imin_imax} gives the SNE characterization similar to Theorem \ref{theorem:char_NE_bidir}. We omit the result in the interest of space, and to avoid repetition.

We now explore the setting when the mutation is from strain $\Lt$ to strain $\Ht$.

\subsection{Mutation from Strain $\Lt$ to $\Ht$}
\label{sec:mut_to_dom_strain}
When the mutation is from strain $\Lt$ to strain $\Ht$, the resulting dynamics is obtained by plugging $\qhl = 0$ in \eqref{eq:bidir_ihdot} and \eqref{eq:bidir_ildot}.

The new dynamics has three potential stationary points for a fixed value of $\zs$. As before, we denote the steady-state values by the tuple $(\Ststar, \Ihstar, \Ilstar)$. We first introduce the following notation:
\begin{align*}
        \mathcal{D}_{2} &:= \betah \gammal - \betal \gammah + \betah \qlh - \betal \qlh,
        \\\mathcal{N}^{\Lt}_{2}(\zs) &:= \gammal + \qlh - \hatbetal(\zs).
    \end{align*}

The three possible equilibria for a given $\zs \in [0, 1]$ are:
\begin{align*}
    \textbf{E}^{\Lt,\Ht}_{1} &= (1, 0, 0);
     \\ \textbf{E}^{\Lt,\Ht}_{2}(\zs) &= \bigg(\frac{\gammah}{\hatbetah(\zs)}, \frac{\hatbetah(\zs) - \gammah}{\hatbetah(\zs)}, 0\bigg);
     \\ \textbf{E}^{\Lt,\Ht}_{3}(\zs) &= \bigg(\frac{\gammal + \qlh}{\hatbetal(\zs)}, 
     \\ & \qquad \frac{\qlh \mathcal{N}^{\Lt}_{2}(\zs)}{w(\zs) \mathcal{D}_{2}}, \frac{-\mathcal{N}^{\Lt}_{2}(\zs) (\mathcal{D}_{2} + \betal \qlh)}{\hatbetal(\zs) \mathcal{D}_{2}} \bigg).
\end{align*}
We now state the following proposition.
\begin{proposition}
    Under the assumption of $\frac{\betah}{\gammah} > \frac{\betal}{\gammal}$, equilibrium $\textbf{E}^{\Lt,\Ht}_{3}(\zs)$ does not exist.
    \label{prop:e3_not_exist}
\end{proposition}
\begin{proof}
    Equilibrium $\textbf{E}^{\Lt,\Ht}_{3}(\zs)$ has steady-state infection states $\Ihstar(\zs) = \frac{\qlh \mathcal{N}^{\Lt}_{2}(\zs)}{w(\zs) \mathcal{D}_{2}}$ and $\Ilstar(\zs) = \frac{-\mathcal{N}^{\Lt}_{2}(\zs) (\mathcal{D}_{2} + \betal \qlh)}{\hatbetal(\zs) \mathcal{D}_{2}}$. Clearly, for $\textbf{E}^{\Lt,\Ht}_{3}(\zs)$ to exist, it is necessary that $\Ihstar(\zs), \Ilstar(\zs) > 0$. We first assume that $\mathcal{D}_{2} > 0$, which implies that $\mathcal{N}^{\Lt}_{2}(\zs) > 0$ must hold for $\Ihstar(\zs)$ to be positive. Since, $\betal \qlh > 0$, the quantity $\Ilstar(\zs)$ is negative which is not possible. 
    
    Similarly, if $\mathcal{D}_{2} < 0$, then we must have $\mathcal{N}^{\Lt}_{2}(\zs) < 0$ to ensure $\Ihstar(\zs) > 0$. For $\Ilstar(\zs)$ to be positive, the following must be true:
    \begin{align*}
        &\mathcal{D}_{2} + \betal \qlh < 0 
        \\ \implies &\betah \gammal - \betal \gammah + \betah \qlh < 0
        \\ \implies &\underbrace{\frac{\betah}{\gammah} - \frac{\betal}{\gammal}}_{> 0} < \underbrace{-\frac{\betah \qlh}{\gammah \gammal}}_{< 0},
    \end{align*}
    which is a contradiction. Thus, proportions $\Ihstar(\zs)$ and $\Ilstar(\zs)$ cannot be simultaneously positive. Therefore, $\textbf{E}^{\Lt,\Ht}_{3}(\zs)$ does not exist.
\end{proof}

Thus, the SIS epidemic model with mutation \eqref{eq:sys_dyn} reduces to a standard single virus model (studied in past works \cite{satapathi2022epidemic}), irrespective of the protection adoption behavior of susceptible agents. Higher reproduction number of strain $\Ht$, and the mutation rates of $\qhl = 0$ and $\qlh > 0$ drive the fraction infected by strain $\Lt$ to become susceptible, or infected by strain $\Ht$. Consequently, it is impossible for strain $\Lt$ to survive at equilibrium, i.e., equilibrium $\textbf{E}^{\Lt,\Ht}_{3}(\zs)$ does not exist. In the following section, we numerically illustrate our theoretical findings as the mutation rates and protection cost vary. 

\section{Numerical Simulations}\label{sec:num_sim}

\begin{figure*}[h!]
\centering
  \subfigure{\includegraphics[width=55mm]{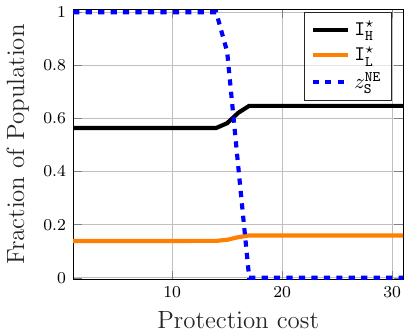}}
  \hspace{4mm}
  \subfigure{\includegraphics[width=55mm]{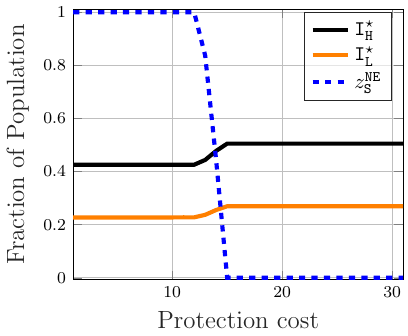}}
  \hspace{4mm}
  \subfigure{\includegraphics[width=55mm]{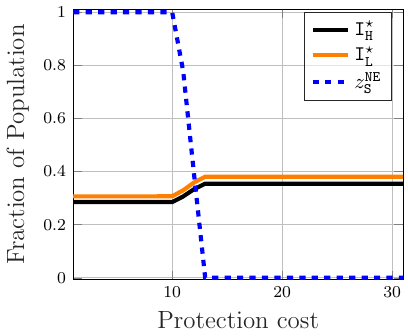}}
  \caption{Variation of steady-state infection levels $(\Ih^{\star}, \Il^{\star})$, and Nash equilibrium $(\zsne)$ with protection cost $(\cp)$, for mutation rates (left) $\qhl = 0.05, \qlh = 0.1$; (middle) $\qhl = 0.1, \qlh = 0.1$; and (right) $\qhl = 0.18, \qlh = 0.1$.}
  \label{fig:bidir}
\end{figure*}

\begin{figure*}[h!]
\centering
  \subfigure{\includegraphics[width=55mm]{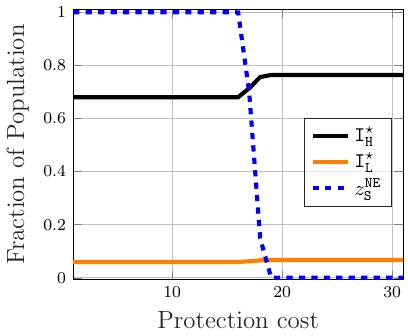}}
  \hspace{4mm}
  \subfigure{\includegraphics[width=55mm]{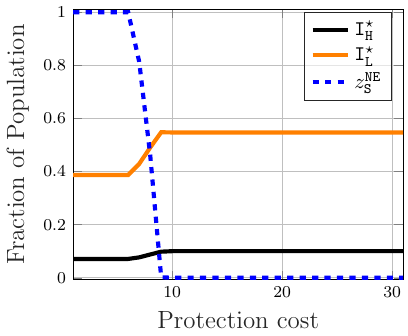}}
  \hspace{4mm}
  \subfigure{\includegraphics[width=55mm]{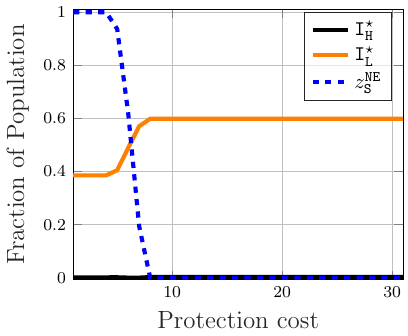}}
  \caption{Variation of steady-state infection levels $(\Ih^{\star}, \Il^{\star})$ and Nash equilibrium $(\zsne)$ with protection cost $(\cp)$, for mutation rate (left) $\qhl = 0.01$; (middle) $\qhl = 0.13$; and (right) $\qhl = 0.19$.}
  \label{fig:var_with_cp}
\end{figure*}

We use the following parameter values in our simulations. The authors in \cite{li2020early} estimated the reproduction number of coronavirus as $2.2$, at the onset of the pandemic. As the virus mutated, some researchers estimated the reproduction number of the new strain to lie in the range of $4.7 \text{ and } 6.6$ (see \cite{sanche2020novel}), resulting in two variants surviving with different reproduction numbers. Accordingly, we select the transmission and recovery rates as given below; the reproduction number of strain $\Lt$ is $2.5$, and that of strain $\Ht$ is $6.5$.

\begin{center}
\begin{tabular}{|c | c | c | c | c | c | c |} 
 \hline
 $\betah$ & $\betal$ & $\gammah$ & $\gammal$ & $\alpha$ & $\rhol$ & $\rhoh$\\ [0.5ex] 
 % & $\St(0)$ & $\Ih(0)$ & $\Il(0)$
 \hline
 0.65 & 0.5 & 0.1 & 0.2 & 0.65 & 70 & 100\\ \hline
 %& 0.7 & 0.2  & 0.1
\end{tabular}
\end{center}

\um{First, we highlight some of our main findings in the general case of bi-directional mutation, followed by validation of some of our findings obtained in the case of mutation from strain $\Ht$ to $\Lt$. In the simulations of bi-directional mutation, we select three cases: (i) $\qhl (= 0.05) < \qlh (= 0.1)$; (ii) $\qhl (= 0.1) = \qlh (= 0.1)$; and (iii) $\qhl (= 0.18) > \qlh (= 0.1)$. The SNE obtained under these parameters for different values of $\cp$ are included in Figure \ref{fig:bidir}.

Initial fractions of the population in different states are chosen as $\St(0) = 0.8, \Ih(0) = 0.1$, and $\Il(0) = 0.1$. We vary the protection cost from $\cp = 1$ to $\cp = 31$. Susceptible agents are assumed to update their Nash strategy based on the Smith dynamics \cite{sandholm2010population}, and we find that the learning dynamics converges to the SNE.  Figure \ref{fig:bidir} shows that at a lower protection cost all susceptible agents adopt protection, with the fraction decreasing as the cost increases, finally with the entire fraction of susceptible agents remaining unprotected at a high cost. Observe that for various mutation rates, both the strains coexist at the steady-state. Furthermore, the infection states $\Ihstar$ and $\Ilstar$ at equilibrium are monotonically increasing with a decrease in $\zsne$. These results are aligned with the findings in Lemma \ref{prop:states_mon_dec}. 

Note that when the mutation rate $\qhl$ is smaller, or comparable to the rate $\qlh$, then the fraction of susceptible agents adopting protection increases. This is because at a lower mutation rate from strain $\Ht$ to $\Lt$, or when both the mutation rates $\qhl$ and $\qlh$ are comparable, there exists a higher fraction of the population who are infected by strain $\Ht$. Since strain $\Ht$ is associated with a larger reproduction number and loss parameter $\rho_{\Ht}$, we find a higher fraction of susceptible individuals to be more inclined to adopt protection. When the mutation rate from strain $\Ht$ to $\Lt$ ($\qhl$) is higher than the rate $\qlh$ (Figure \ref{fig:bidir}, right), then the fraction of the population infected by strain $\Lt$ is higher. Notice an analogous phenomenon in the left panel of Figure \ref{fig:bidir}, when $\qlh > \qhl$ and the fraction infected by strain $\Ht$ is higher. When both mutation rates are equal, the relation $\Ihstar > \Ilstar$ (Figure \ref{fig:bidir}, middle) is observed due to a higher transmission rate, and reproduction number of strain $\Ht$.}

\um{Finally, we validate our findings obtained in uni-directional mutation model, when the mutation is from strain $\Ht$ to $\Lt$.} The mutation rate threshold for the above set of parameters is $\frac{\betah \gammal}{\betal} - \gammah = 0.16$ (as stated in Corollary \ref{cor:GAS_qlh=0}). Initial fractions of the population remain same. We select three different mutation rates and demonstrate the variation of infected proportions $(\Ih^{\star}, \Il^{\star})$ and the proportion that adopt protection $(\zsne)$ at the SNE with protection cost. 

\um{Similar to Figure \ref{fig:bidir},} Figure \ref{fig:var_with_cp} shows that when protection cost is low, all susceptible agents adopt protection, while at larger values of $\cp$, remaining unprotected is the Nash strategy. For the first case, we choose a very low mutation rate of $\qhl = 0.01$, i.e., the rate of transition from strain $\Ht$ to strain $\Lt$ is extremely small. This is illustrated by the plot in the left panel of Figure \ref{fig:var_with_cp}. As expected, we observe that a large proportion of the population remains infected by strain $\Ht$, while a small fraction is infected by strain $\Lt$. Since $\qhl = 0.01 < 0.16$, we observe existence of the stable equilibrium of coexisting viruses $\textbf{E}^{\Ht, \Lt}_{3}$. When $\qhl = 0.13$, i.e., from infection state $\Ih$ to state $\Il$ is somewhat larger, the middle panel of the figure shows that $\Ih^{\star}$ has reduced considerably compared to the previous case, and $\Il^{\star}$ is now higher than $\Ih^{\star}$. Since $\qhl = 0.13 < 0.16$ still holds, we observe a non-zero fraction of the population infected by strain $\Ht$. Further increasing the mutation rate to $\qhl = 0.19$ (right panel), we depict convergence to the stable single strain equilibrium $\textbf{E}^{\Ht, \Lt}_{2}$. These observations validate our theoretical results.

On examining the three plots, we again observe a shift in the behavior of susceptible agents, i.e., for low mutation rates, susceptible agents choose protection even when the cost of protection adoption is sufficiently large. As the mutation rate increases, agents are reluctant to adopt protection even for smaller values of $\cp$. \um{This shift is similar to the protection adoption behavior depicted in Figure \ref{fig:bidir}, and it} arises since the dominant strain $\Ht$ has a larger infection rate, reproduction number, and a larger value of loss parameter $\rho_{\Ht}$. Note that with increasing mutation rate $\qhl$, the proportion $\Ih^{\star}$ decreases, and finally reduces to zero. Hence, even though the proportion $\Il^{\star}$ increases, the effective loss of becoming infected falls, resulting in fewer agents adopting protection. 

\section{Conclusion}\label{sec:conc}
\um{We investigated the impact of game-theoretic protection adoption on infection prevalence in a susceptible-infected-susceptible (SIS) epidemic model involving two mutant strains. Our analysis encompassed both the general case of bi-directional mutation and the special case of uni-directional mutation. For each setting, we first established the existence and stability of equilibrium points for a fixed level of protection adoption by susceptible agents. When mutation is bi-directional, or when it occurs from the stronger strain to the weaker one, we demonstrated the possibility of coexistence of both strains at equilibrium and characterized the stationary Nash equilibrium (SNE) of the underlying game. Our findings highlighted the influence of mutation on infection prevalence and protection strategies. For example, we showed that higher mutation rates to the dominant strain lead to higher protection adoption among individuals, whereas a lower rate reduces the incentive for protection.

This work opens several avenues for future research. One potential direction is to extend the analysis to networked epidemic models with heterogeneous node degrees. In this work, we have assumed that the mutation rates are constant. A possible direction is to model the mutation rates as functions of the protection adoption behavior of the susceptible agents. Finally, the current study assumes equal effectiveness of protection against both strains; relaxing this assumption to reflect more realistic scenarios presents an interesting area for further investigation.}

\appendix

\section{\underline{\textbf{Proof of Theorem \ref{thm:existence}}}}
\label{sec:appen_existence}

\begin{proof}
Recall from Remark \ref{remark:ee_not_exist} that endemic equilibrium does not exist when reproduction number satisfies the relation $\mathcal{R}_{0} \leq 1$. Furthermore, when Assumption \ref{assump:mutation_rates} holds, we have $\mathcal{R}_{0} > 1$. We prove the proposition by leveraging \emph{Index Theory} \cite{strogatz2024nonlinear}. To this end we define a closed curve, AEBFCGHA, as shown in Figure \ref{fig:closed_curve}.
    
\begin{figure}[h!]
\centering
  \subfigure{\includegraphics[width=0.7\linewidth]{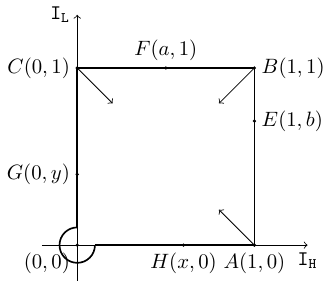}}
  \caption{Figure depicting a closed curve, and the associated vector fields of dynamics \eqref{eq:bidir_ihdot} and \eqref{eq:bidir_ildot}.}
  \label{fig:closed_curve}
\end{figure}
We select a closed geometrical shape similar to a square, with one of the the corners not touching the origin. Recall that the origin of $(\Ih, \Il) = (0, 0)$ in our system denotes the disease-free equilibrium. Points $A = (1, 0), B = (1, 1)$ and $C = (0, 1)$ denote the three corners of the closed curve, whereas instead of the fourth corner there exists a detour which includes the disease-free equilibrium. The detour around the origin is an infinitesimally small three-quarter circle with its center at the origin.

We now compute the vector fields $\dot{\textbf{x}} := (\dot{\mathtt{I}}_{\mathtt{H}}, \dot{\mathtt{I}}_{\mathtt{L}})$ at points $A, B$ and $C$ as:
\begin{align}
    &\dot{\textbf{x}}(A) = \begin{bmatrix}
        \dot{\mathtt{I}}_{\mathtt{H}}(A)\\
        \dot{\mathtt{I}}_{\mathtt{L}}(A)
    \end{bmatrix} = \begin{bmatrix}
        -(\qhl + \gammah) \\
        \qhl
    \end{bmatrix}, \label{eq:pointA}
    \\ &\dot{\textbf{x}}(B) = \begin{bmatrix}
        \dot{\mathtt{I}}_{\mathtt{H}}(B)\\
        \dot{\mathtt{I}}_{\mathtt{L}}(B)
    \end{bmatrix} = \begin{bmatrix}
        -\hatbetah + \qlh -(\qhl + \gammah) \\
        -\hatbetal + \qhl - (\qlh + \gammal)
    \end{bmatrix}, \label{eq:pointB}
    \\ &\dot{\textbf{x}}(C) = \begin{bmatrix}
        \dot{\mathtt{I}}_{\mathtt{H}}(C)\\
        \dot{\mathtt{I}}_{\mathtt{L}}(C)
    \end{bmatrix} = \begin{bmatrix}
        \qlh \\
        -(\qlh + \gammal)
    \end{bmatrix}. \label{eq:pointC}
\end{align}
Now, we examine directions of the vector field at these points to determine its angular variation, as the closed curve is traversed anti-clockwise. Note from \eqref{eq:pointA} that the first component of $\dot{\textbf{x}}(A)$ is negative, whereas its second component is positive for all parameter values. Thus, we obtain the direction of the vector field at $A$ as is depicted in Figure \ref{fig:closed_curve}. Under Assumption \ref{assump:mutation_rates}, both components of $\dot{\textbf{x}}(B)$ are negative. Similarly, the first component of $\dot{\textbf{x}}(C)$ is positive, while the second component is negative. Hence, we obtain the vector field directions at $B$ and $C$ as is shown in Figure \ref{fig:closed_curve}. 

Now, we consider the vector field direction along the three-quarter circular arc $D_{1} D D_{2}$ in Figure \ref{fig:quar_cir1}. Observe that an arbitrary point $D(R, \phi)$ on the three-quarter circle can be represented by the radius $R$, and the angle $\phi$. Note that the radius $R > 0$ is infinitesimally small, and the figure is the magnified representation. The states are denoted by $$\Ih = R \cos \phi, \;\; \Il = R \sin \phi,$$ where $\phi$ varies from $\frac{\pi}{2}$ at point $D_{1}$ to $2 \pi$ at $D_{2}$. Consequently, the expression of the vector field at a point $D(R, \phi)$ is obtained by substituting $\Ih$ and $\Il$ in \eqref{eq:bidir_ihdot} and \eqref{eq:bidir_ildot}, under the approximation of $R \rightarrow 0 \Rightarrow R^2 \cong 0$, and is given by:
\begin{align}
    \dot{\textbf{x}}(D) &= \begin{bmatrix}
        \big( \hatbetah - (\qhl + \gammah) \big) R \cos \phi + \qlh R \sin \phi\\
        \big( \hatbetal - (\qlh + \gammal) \big) R \sin \phi + \qhl R \cos \phi
    \end{bmatrix}.
    \label{eq:gen_exp_R_phi}
\end{align}
When $\phi = \frac{\pi}{2}$ holds, the vector field reduces to $$\dot{\textbf{x}} (D_{1}) = \begin{bmatrix}
    \qlh R\\
    \hatbetal R - (\qlh + \gammal) R
\end{bmatrix},$$ and at $\phi = 2 \pi$, we obtain $$\dot{\textbf{x}} (D_{2}) = \begin{bmatrix}
    \hatbetah R - (\qhl + \gammah) R\\
    \qhl R
\end{bmatrix}.$$
Let $\theta$ be the angle of the vector field $\dot{\textbf{x}}(\cdot)$ (in the anti-clockwise direction) with respect to the horizontal axis $\Ih$, as shown in Figure \ref{fig:quar_cir1}. We now analyze the variation of $\theta$ with $\phi$. In other words, our objective is to determine the direction in which the vector field rotates, as one traverses in anti-clockwise direction from $D_{1}$ to $D_{2}$. The general expression of $\theta$ is obtained from \eqref{eq:gen_exp_R_phi} as 
\begin{align*}
    &\tan \theta = \frac{\dot{\mathtt{I}}_{\mathtt{L}}}{\dot{\mathtt{I}}_{\mathtt{H}}}
    \\\Rightarrow &\theta = \tan^{-1} \! \bigg(\frac{\big(\hatbetal - (\qlh + \gammal) \big) \sin \phi + \qhl \cos \phi}{\big(\hatbetah - (\qhl + \gammah) \big) \cos \phi + \qlh \sin \phi} \bigg).
\end{align*}
To find the variation of $\theta$ with $\phi$, we differentiate it to obtain
\begin{align*}
    \frac{d \theta}{d \phi} =& \frac{1}{1 + \bigg(\frac{\big(\hatbetal - (\qlh + \gammal) \big) \sin \phi + \qhl \cos \phi}{\big(\hatbetah - (\qhl + \gammah) \big) \cos \phi + \qlh \sin \phi} \bigg)^{2}} 
    \\ &\times \frac{d}{d \phi} \bigg(\frac{\big(\hatbetal - (\qlh + \gammal) \big) \sin \phi + \qhl \cos \phi}{\big(\hatbetah - (\qhl + \gammah) \big) \cos \phi + \qlh \sin \phi} \bigg),
\end{align*}
which on further simplifications yields
\begin{align}
    &\frac{d \theta}{d \phi} = \frac{\mathcal{N}}{\mathcal{D}(\phi)},
    \label{eq:d_theta_by_d_phi}
\end{align}
where 
\begin{align*}
    \mathcal{N} &:= \big( \hatbetal - (\qlh + \gammal) \big) \big( \hatbetah - (\qhl + \gammah) \big) - \qlh \qhl,
    \\\mathcal{D}(\phi) &:= \big( \big( \hatbetal - (\qlh + \gammal) \big) \sin \phi + \qhl \cos \phi \big)^{2} 
    \\ &+ \big( \big( \hatbetah - (\qhl + \gammah) \big) \cos \phi + \qlh \sin \phi \big)^{2}.
\end{align*}

\begin{figure}[t!]
\centering
  \subfigure{\includegraphics[width=0.7\linewidth]{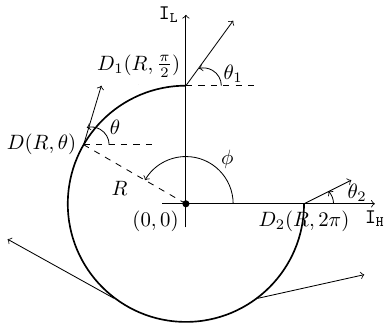}}
  \caption{Three-quarter circular arc (magnified) with center at origin and both eigenvalues being positive, i.e., $\theta_{1} > \theta_{2}$.}
  \label{fig:quar_cir1}
\end{figure}

Now, under Assumption \ref{assump:mutation_rates}, both $\hatbetal - (\qlh + \gammal) > 0$, and $\hatbetah - (\qhl + \gammah) > 0$ are satisfied, which implies that $$\hatbetal + \hatbetah - (\qlh + \gammal + \qhl + \gammah) > 0,$$ holds true. Now, two cases may arise under the constraint of $\mathcal{R}_{0} > 1$, depending upon the characteristics of the eigenvalues. 

\vspace{1mm}
\textbf{Case 1: Both eigenvalues of $\mathcal{J}(\textbf{E}_{\mathtt{DFE}})$ are positive.}
\vspace{1mm}

Both eigenvalues being positive also implies that the determinant of $\mathcal{J}(\textbf{E}_{\mathtt{DFE}})$ in \eqref{eq:jac_dfe} is positive, i.e.,
\begin{align*}
    &\big(\hatbetah - (\qhl + \gammah) \big) \big(\hatbetal - (\qlh + \gammal) \big) - \qlh \qhl > 0
    \\ \Rightarrow &\frac{\hatbetal - (\qlh + \gammal)}{\qlh} > \frac{\qhl}{\hatbetah - (\qhl + \gammah)}
    \\ \Rightarrow &\tan^{-1} \!\! \bigg(\frac{\hatbetal - (\qlh + \gammal)}{\qlh}\bigg) \!\! > \!\! \tan^{-1} \!\! \bigg(\frac{\qhl}{\hatbetah - (\qhl + \gammah)} \bigg)
    \\ \Rightarrow &\theta_{1} > \theta_{2},
\end{align*}
implying that the angle of the vector field at $D_{1}(R, \frac{\pi}{2})$ is higher than the angle at $D_{2}(R, 2 \pi)$ (see Figure \ref{fig:quar_cir1}). In addition, observe that $\frac{d \theta}{d \phi} > 0$ (equation \eqref{eq:d_theta_by_d_phi}) in this case, which describes the angular variation of the vector field, i.e., the vector field rotates in an anti-clockwise direction throughout as $\phi$ increases from $\frac{\pi}{2}$ to $2 \pi$.

We now compute the total angle covered by the vector field when the closed curve is traversed for one complete anti-clockwise rotation. First, we start with point $D_{2}$ (refer to Figures \ref{fig:closed_curve} and \ref{fig:quar_cir1}), then traversing anti-clockwise, we reach $A$, followed by points $B$, $C$ and $D_{1}$, covering an angle equal to exactly $2\pi + \theta_{1} - \theta_{2}$. As we traverse from point $D_{1}$ to $D_{2}$, the vector field continues to rotate in an anti-clockwise direction, and consequently, reaches $D_{2}$, covering an angle of $2\pi + \theta_{2} - \theta_{1}$. Therefore, a complete traversal of the closed curve in anti-clockwise direction covers a total angle of exactly $4\pi$. Note that when both the eigenvalues are positive the disease-free equilibrium is a source, and the vector field is never directed inward pointing towards the origin. This excludes the possibility of an anti-clockwise angular rotation of the vector field by greater than $2 \pi$ while traversing the three-quarter circular arc.

It remains is to rule out the possibility of a complete angular variation by a multiple of $2\pi$ in between the corner points, i.e., in between $A$ and $B$; between $B$ and $C$ and so on. In order to do so, we choose four intermediate random points $E, F, G$ and $H$ (Figure \ref{fig:closed_curve}), and analyze direction of the vector field at these points. The vector field at point $E(1, b)$ is computed as:
\begin{equation}
    \dot{\textbf{x}}(E) = \begin{bmatrix}
        \hatbetah b + \qlh b \! - \! (\qhl + \gammah)\\
        -\hatbetal b^2 \! + \! \qhl \! - \! (\qlh + \gammal) b
    \end{bmatrix}.
    \label{eq:pointE}
\end{equation}
The first component is negative under Assumption \ref{assump:mutation_rates}. When $b$ is small and $b \rightarrow 0$, then $-\hatbetal b^2 \! + \! \qhl \! - \! (\qlh + \gammal) b \rightarrow \qhl$, i.e., the second component is positive. As $b$ increases, $-\hatbetal b^2 \! + \! \qhl \! - \! (\qlh + \gammal) b$ monotonically decreases, and as $b \rightarrow 1$, the second component $-\hatbetal b^2 \! + \! \qhl \! - \! (\qlh + \gammal) b \rightarrow -\hatbetal \! + \! \qhl \! - \! (\qlh + \gammal)$, which is negative under the same assumption. Thus, angle of $\dot{\textbf{x}}$ monotonically changes from the direction at point $A(1, 0)$ to $B(1, 1)$ (see Figure \ref{fig:closed_curve}), without a complete rotation by a multiple of $2\pi$ in between.

Similar analysis for arbitrary points $F(a, 1), G(0, y)$ and $H(x, 0)$ lead to the same observation of the absence of a complete rotation of the vector field by a multiple of $2\pi$ in between the points.

Thus, the index $I$, of the closed curve with respect to the vector field is computed as $$I = \frac{1}{2\pi} \times 4\pi = +2.$$ Note that by Lemma \ref{lemm:stab_ee}, an endemic equilibrium is locally stable when it exists. This rules out the possibility of existence of an unstable saddle point (which carries an index of $-1$) as an endemic equilibrium. In addition, since both eigenvalues are positive, the disease-free equilibrium is a source and therefore carries an index of $+1$. Thus, the index of the closed curve, $I = +2$ implies the existence of exactly one endemic equilibrium, which establishes the uniqueness of the locally stable endemic equilibrium.

\vspace{1mm}
\textbf{Case 2: Exactly one of the eigenvalues of $\mathcal{J}(\textbf{E}_{\mathtt{DFE}})$ is negative.}
\vspace{1mm}

Presence of a negative and positive eigenvalue implies that the determinant of $\mathcal{J}(\textbf{E}_{\mathtt{DFE}})$ in \eqref{eq:jac_dfe} is negative, i.e.,
\begin{align*}
    &\big(\hatbetah - (\qhl + \gammah) \big) \big(\hatbetal - (\qlh + \gammal) \big) - \qlh \qhl < 0
    \\ \Rightarrow &\frac{\hatbetal - (\qlh + \gammal)}{\qlh} < \frac{\qhl}{\hatbetah - (\qhl + \gammah)}
    \\ \Rightarrow & \tan^{-1} \bigg(\frac{\hatbetal - (\qlh + \gammal)}{\qlh}\bigg) < \tan^{-1} \bigg(\frac{\qhl}{\hatbetah - (\qhl + \gammah)} \bigg)
    \\ \Rightarrow &\theta_{1} < \theta_{2},
\end{align*}
implying that the angle of the vector field at $D_{1}(R, \frac{\pi}{2})$ is smaller than the angle at $D_{2}(R, 2 \pi)$. Furthermore, from \eqref{eq:d_theta_by_d_phi} we observe that $\frac{d \theta}{d \phi} < 0$ in this case, i.e., the vector field rotates in a clockwise direction throughout as $\phi$ increases from $\frac{\pi}{2}$ to $2 \pi$, as shown in Figure \ref{fig:Full_vector_fld_ev_negative}.

\begin{figure}[t!]
\centering
  \subfigure{\includegraphics[width=0.7\linewidth]{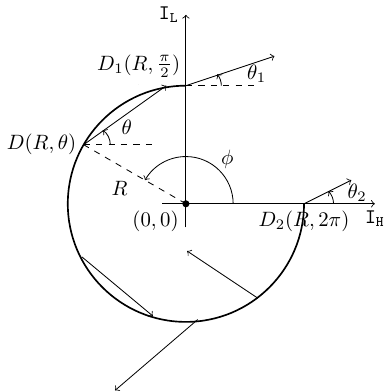}}
  \caption{Three-quarter circular arc (highly magnified) with center at origin and exactly one positive eigenvalue, i.e., $\theta_{1} < \theta_{2}$.}
  \label{fig:Full_vector_fld_ev_negative}
\end{figure}

Now, for computing the total angle covered when the closed curve is traversed for one complete rotation in an anti-clockwise direction, we start with point $D_{2}$ (refer to Figures \ref{fig:closed_curve} and \ref{fig:Full_vector_fld_ev_negative}), then traversing anti-clockwise, we reach $A$, followed by points $B$, $C$ and $D_{1}$, covering an angle of $2 \pi + \theta_{1} - \theta_{2}$. Since, one of the eigenvalues is negative, the origin behaves as a saddle, and at some point the vector field must be in an inward direction, towards from the origin. Hence, on traversing point $D_{2}$ from $D_{1}$ the vector field rotates clockwise, covering an angle of $-2 \pi + \theta_{2} - \theta_{1}$. Therefore, a complete anti-clockwise traversal of the curve covers a net angle of $0$.

Similar to Case 1, it can be shown that a complete rotation of the vector field by a multiple of $2\pi$ in between the points $A$ and $B$, $B$ and $C$ etc. is absent. Finally, we rule out the possibility of clockwise variation of the vector field by an angle smaller than $-2 \pi$. Observe that when the vector field rotates by an angle smaller than $-2 \pi$ while traversing the arc, it implies that index must satisfy $I \leq -1$. The equality $I = -1$ indicates that the disease-free equilibrium is the only equilibrium that exists. However, in the absence of closed orbits (as shown in Theorem \ref{thm:stab_ee}), the trajectories fail to converge, since the origin is unstable. Thus, $I = -1$ is impossible. Now, when $I < -1$ is satisfied, it implies that at least one more unstable saddle exists as the endemic equilibrium. This is also not possible, since it violates Lemma \ref{lemm:stab_ee}. Thus, the index $I$, in this case is computed as $$I = \frac{1}{2\pi} \times 0 = 0.$$ Since exactly one eigenvalue is negative, the disease-free equilibrium is a saddle and therefore carries an index of $-1$. Non-existence of an unstable saddle implies the existence of exactly one endemic equilibrium (with index as $+1$, that cancels out the index of the disease-free equilibrium), which establishes the uniqueness of the locally stable endemic equilibrium when one of the eigenvalues is positive.

We conclude the proof by stating that $\mathcal{R}_{0} > 1$ is necessary for existence of the endemic equilibrium, and this follows from the discussion in Remark \ref{remark:ee_not_exist}.
\end{proof}

\section{\underline{\textbf{Proof of Lemma \ref{prop:states_mon_dec}}}}
\label{sec:appen_states_mon_dec}

\begin{proof}
    We begin the proof by rewriting \eqref{eq:bidir_dot_ih_il=0} at a given $\zs \in [0, 1]$ as 
    \begin{align}
        \frac{\Ilstar(\zs)}{\Ihstar(\zs)} &= \frac{\gammah + \qhl - \hatbetah(\zs) \Ststar(\zs)}{\qlh} \nonumber \\ &=\frac{\qhl}{\gammal + \qlh - \hatbetal(\zs) \Ststar(\zs)}.
        \label{eq:il_by_ih}
    \end{align}
     Let $z_{\mathtt{S}_{2}} > z_{\mathtt{S}_{1}}$ be two protection strategies adopted by the susceptible agents. For the equality in \eqref{eq:il_by_ih} to be preserved with an increase in $\zs$, it is necessary that when protection strategy increases from $z_{\mathtt{S}_{1}}$ to $z_{\mathtt{S}_{2}}$, either both quantities $\frac{\gammah + \qhl - \hatbetah(z_{\mathtt{S}_{1}}) \Ststar(z_{\mathtt{S}_{1}})}{\qlh}$ and $\frac{\qhl}{\gammal + \qlh - \hatbetal(z_{\mathtt{S}_{1}}) \Ststar(z_{\mathtt{S}_{1}})}$, simultaneously increase, or both quantities simultaneously decrease, or both remain constant. First, we hypothesize that both quantities increase as $\zs$ is increased from $z_{\mathtt{S}_{1}}$ to $z_{\mathtt{S}_{2}}$. This is true when $\gammah + \qhl - \hatbetah(\zs) \Ststar(\zs)$ increases with an increased $\zs$, i.e., the relation $$\hatbetah(z_{\mathtt{S}_{1}}) \Ststar(z_{\mathtt{S}_{1}}) > \hatbetah(z_{\mathtt{S}_{2}}) \Ststar(z_{\mathtt{S}_{2}}),$$ must hold, which implies
    \begin{equation}
        \frac{\alpha z_{\mathtt{S}_{1}} + 1 - z_{\mathtt{S}_{1}}}{\alpha z_{\mathtt{S}_{2}} + 1 - z_{\mathtt{S}_{2}}} > \frac{\Ststar(z_{\mathtt{S}_{2}})}{\Ststar(z_{\mathtt{S}_{1}})}.
        \label{eq:fractions}
    \end{equation}
    Similarly, the other quantity increases when $\gammal + \qlh - \hatbetal(\zs) \Ststar(\zs)$ decreases when $\zs$ is increased from $z_{\mathtt{S}_{1}}$ to $z_{\mathtt{S}_{2}}$, i.e., when
    \begin{align*}
        &\hatbetal(z_{\mathtt{S}_{1}}) \Ststar(z_{\mathtt{S}_{1}}) < \hatbetal(z_{\mathtt{S}_{2}}) \Ststar(z_{\mathtt{S}_{2}})
        \\ \implies &\frac{\alpha z_{\mathtt{S}_{1}} + 1 - z_{\mathtt{S}_{1}}}{\alpha z_{\mathtt{S}_{2}} + 1 - z_{\mathtt{S}_{2}}} < \frac{\Ststar(z_{\mathtt{S}_{2}})}{\Ststar(z_{\mathtt{S}_{1}})},
    \end{align*}
    holds. Note that the above relation is in contradiction to the inequality obtained in \eqref{eq:fractions}. This proves that the two quantities of interest can not increase simultaneously. Proceeding along similar lines, it can be shown that it is impossible for the two quantities to decrease simultaneously. Thus, the only possibility is that there exists a constant $C$, such that
    \begin{align*}
        &\frac{\gammah + \qhl - \hatbetah(\zs) \Ststar(\zs)}{\qlh} 
        \\ &= \frac{\qhl}{\gammal + \qlh - \hatbetal(\zs) \Ststar(\zs)} = C
        \\ \Rightarrow & \frac{\Ilstar(\zs)}{\Ihstar(\zs)} = C,
    \end{align*}
    holds for all $\zs$.
    
    Note that both $\hatbetah(\zs)$ and $\hatbetal(\zs)$ decrease monotonically as $\zs$ increases from $0$ to $1$. Consequently, for the quantities $\hatbetah(\zs) \Ststar(\zs)$ and $\hatbetal(\zs) \Ststar(\zs)$ to remain constant for all $\zs$, the state $\Ststar(\zs)$ must be monotonically increasing in $\zs$, unless $\Ststar(\zs) = 0$ identically holds for all $\zs \in [0, 1]$. It is easy to see that substituting $\Ststar(\zs) = 0$ in \eqref{eq:sys_dyn} (suppressing the dependency on $\zs$), implies,
    \begin{equation*}
        \dot{\St} = -\hatbetah \Ihstar \Ststar - \hatbetal \Ilstar \Ststar + \gammah \Ihstar + \gammal \Ilstar = 0,
    \end{equation*}
    resulting in $\gammah \Ihstar = - \gammal \Ilstar$, which is not possible. Thus, \eqref{eq:il_by_ih} is preserved only when $\Ststar(\zs)$ monotonically increases with $\zs$. This in turn implies that sum of the infection states, i.e., $\Ihstar(\zs) + \Ilstar(\zs)$, monotonically decreases with an increase in $\zs$. Exploiting this result, we conclude that $\Ihstar(\zs)$ monotonically decreases with an increase in $\zs$, since $\Ihstar(\zs) + \Ilstar(\zs) = \Ihstar(\zs) + C \Ihstar(\zs)$. Similarly, $\Ilstar(\zs)$ also monotonically decreases with $\zs$. This concludes our proof.
\end{proof}

\section{\underline{\textbf{Proof of Corollary \ref{cor:GAS_qlh=0}}}}
\label{sec:appen_corollary}

\begin{proof}
    % \textbf{Case 2:} Single-strain equilibrium, $\textbf{E}^{\Ht,\Lt}_{2}(\zs)$
    %     \vspace{1mm}
        Note that Assumption \ref{assump:mutation_rates} is not violated in the limiting case of $\qlh = 0$. GAS of $\textbf{E}^{\Ht,\Lt}_{1}$ and $\textbf{E}^{\Ht,\Lt}_{3}(\zs)$ follows directly from Theorem \ref{thm:stab_dfe}, and Lemma \ref{lemm:stab_ee} and Theorem \ref{thm:stab_ee}, with the additional condition of $\qhl < \frac{\betah \gammal}{\betal} - \gammah$ in the case of endemic equilibrium.
        
        For a given $\zs \in [0, 1]$, existence of $\textbf{E}^{\Ht,\Lt}_{2}(\zs)$ is determined by the proportion $\Ilstar$. The condition for its existence is $0 < \Ilstar := \frac{\mathcal{N}^{\Lt}_{1}(\zs)}{\hatbetal(\zs)} = 1 -\frac{\gammal}{\hatbetal(\zs)} < 1$, or equivalently, $\frac{\hatbetal(\zs)}{\gammal} > 1$. 
        The Jacobian matrix at endemic equilibrium $\textbf{E}^{\Ht,\Lt}_{2}(\zs)$ reduces to
        \begin{align*}
            &\mathcal{J}(\textbf{E}^{\Ht,\Lt}_{2}) := 
            \\ &\begin{bmatrix}
            \hatbetah(\zs) \!-\! \hatbetah(\zs) \Ilstar \!-\! \gammah \!-\! \qhl & 0
            \\ -\!\hatbetal(\zs) \Ilstar \!-\! \qhl & \hatbetal(\zs) \!-\! 2 \hatbetal(\zs) \Ilstar - \gammal
        \end{bmatrix},
        \end{align*}
        whose eigenvalues are $\hatbetah(\zs) (1 - \Ilstar) - \gammah - \qhl$ and $\hatbetal(\zs) - 2 \hatbetal(\zs) \Ilstar - \gammal$. The former eigenvalue, on substitution from \eqref{eq:e_3} simplifies to $\frac{\betah \gammal}{\betal} - \qhl - \gammah$. The latter reduces to $\gammal - \hatbetal(\zs)$. To guarantee local stability, the real parts of both the eigenvalues should be negative, which yields the conditions of $\qhl > \frac{\betah \gammal}{\betal} - \gammah$, and $\frac{\hatbetal(\zs)}{\gammal} > 1$. Since the GAS of $\textbf{E}^{\Ht,\Lt}_{1}$ and $\textbf{E}^{\Ht,\Lt}_{3}(\zs)$ are ensured by $\max\bigg(\frac{\hatbetah(\zs)}{\gammah + \qhl}, \frac{\hatbetal(\zs)}{\gammal} \bigg) < 1$, and $\frac{\hatbetah(\zs)}{\gammah + \qhl} > 1$ and $\qhl < \frac{\betah \gammal}{\betal} - \gammah$, respectively, the above derived conditions of $\frac{\hatbetal(\zs)}{\gammal} > 1$ and $\qhl > \frac{\betah \gammal}{\betal} - \gammah$ also guarantee the uniqueness of equilibrium $\textbf{E}^{\Ht,\Lt}_{2}(\zs)$. 
        
        We now establish GAS of $\textbf{E}^{\Ht,\Lt}_{2}(\zs)$ when it exists. When $\qlh$ reduces to zero, Assumption \ref{assump:mutation_rates} and the proof of Theorem \ref{thm:stab_ee} remain valid. Consequently, by Dulac's Criterion we rule out the existence of closed orbits lying entirely in the interior of $(0, 1)^2$. Now, at the boundary of $\Ih = 0$, we observe $\dot{\mathtt{I}}_{\mathtt{H}} = 0$ holds, implying that $\Ih = 0$ is invariant. Thus, the dynamics reduces to $$\dot{\mathtt{I}}_{\mathtt{L}} = \hatbetal (1 - \Il) \Il - \gammal \Il,$$
    i.e., \eqref{eq:mut_h_to_l} reduces to a one-dimensional system, which in turn rules out the possibility of closed orbits at the boundary $\Ih = 0$. Since $\textbf{E}^{\Ht,\Lt}_{2}(\zs)$ is the only locally stable equilibrium under $\qhl > \frac{\betah \gammal}{\betal} - \gammah$ and $\frac{\hatbetal(\zs)}{\gammal} > 1$, in the absence of any closed orbits, all trajectories with initial conditions in $(0, 1)^2$ of the two-dimensional system converge to $\textbf{E}^{\Ht,\Lt}_{2}(\zs)$, i.e., the equilibrium is GAS. This completes our proof.
\end{proof}

\bibliographystyle{plain}        % Include this if you use bibtex 
\bibliography{main_mutation_game}           % and a bib file to produce the 

\begin{thebibliography}{10}

\bibitem{blanchini2008set}
Franco Blanchini and Stefano Miani.
\newblock {\em Set-theoretic Methods in Control}.
\newblock Springer, 2 edition, 2015.

\bibitem{bonhoeffer1994mutation}
Sebastian Bonhoeffer and Martin~A Nowak.
\newblock Mutation and the evolution of virulence.
\newblock {\em Proceedings of the Royal Society of London. Series B: Biological
  Sciences}, 258(1352):133--140, 1994.

\bibitem{bressan2007introduction}
Alberto Bressan and Benedetto Piccoli.
\newblock {\em Introduction to the Mathematical Theory of Control}, volume~1.
\newblock American Institute of Mathematical Sciences Springfield, 2007.

\bibitem{chavda2022delta}
Vivek~P Chavda, Rajashri Bezbaruah, Kangkan Deka, Lawandashisha Nongrang, and
  Tutumoni Kalita.
\newblock The delta and omicron variants of {SARS-C}o{V}-2: {W}hat we know so
  far.
\newblock {\em Vaccines}, 10(11):1926, 2022.

\bibitem{horn2012matrix}
Roger~A Horn and Charles~R Johnson.
\newblock {\em Matrix analysis}.
\newblock Cambridge university press, 2012.

\bibitem{hota2023learning}
Ashish~R Hota, Urmee Maitra, Ezzat Elokda, and Saverio Bolognani.
\newblock Learning to mitigate epidemic risks: A dynamic population game
  approach.
\newblock {\em Dynamic Games and Applications}, 13(4):1106--1129, 2023.

\bibitem{huang2022game}
Yunhan Huang and Quanyan Zhu.
\newblock Game-theoretic frameworks for epidemic spreading and human
  decision-making: A review.
\newblock {\em Dynamic Games and Applications}, 12(1):7--48, 2022.

\bibitem{janson2020networked}
Axel Janson, Sebin Gracy, Philip~E Par{\'e}, Henrik Sandberg, and Karl~H
  Johansson.
\newblock Networked multi-virus spread with a shared resource: Analysis and
  mitigation strategies.
\newblock {\em arXiv preprint arXiv:2011.07569}, 2020.

\bibitem{khanafer2016stability}
Ali Khanafer, Tamer Ba{\c{s}}ar, and Bahman Gharesifard.
\newblock Stability of epidemic models over directed graphs: A positive systems
  approach.
\newblock {\em Automatica}, 74:126--134, 2016.

\bibitem{li2020early}
Qun Li, Xuhua Guan, Peng Wu, Xiaoye Wang, Lei Zhou, Yeqing Tong, Ruiqi Ren,
  Kathy~SM Leung, Eric~HY Lau, Jessica~Y Wong, et~al.
\newblock Early transmission dynamics in {W}uhan, {C}hina, of novel
  coronavirus--infected pneumonia.
\newblock {\em New England journal of medicine}, 382(13):1199--1207, 2020.

\bibitem{liu2019analysis}
Ji~Liu, Philip~E Par{\'e}, Angelia Nedi{\'c}, Choon~Yik Tang, Carolyn~L Beck,
  and Tamer Ba{\c{s}}ar.
\newblock Analysis and control of a continuous-time bi-virus model.
\newblock {\em IEEE Transactions on Automatic Control}, 64(12):4891--4906,
  2019.

\bibitem{10750281}
Urmee Maitra, Ashish~R. Hota, and Philip~E. Paré.
\newblock Optimal bayesian persuasion for containing {SIS} epidemics.
\newblock {\em IEEE Control Systems Letters}, 8:2499--2504, 2024.

\bibitem{maitra2023sis}
Urmee Maitra, Ashish~R Hota, and Vaibhav Srivastava.
\newblock {SIS} epidemic propagation under strategic non-myopic protection: A
  dynamic population game approach.
\newblock {\em IEEE Control Systems Letters}, 7:1578--1583, 2023.

\bibitem{martins2023epidemic}
Nuno~C Martins, Jair Cert{\'o}rio, and Richard~J La.
\newblock Epidemic population games and evolutionary dynamics.
\newblock {\em Automatica}, 153:111016, 2023.

\bibitem{nowak1991evolution}
Martin Nowak.
\newblock The evolution of viruses. competition between horizontal and vertical
  transmission of mobile genes.
\newblock {\em Journal of Theoretical Biology}, 150(3):339--347, 1991.

\bibitem{paarporn2023sis}
Keith Paarporn and Ceyhun Eksin.
\newblock {SIS} epidemics coupled with evolutionary social distancing dynamics.
\newblock In {\em American Control Conference}, pages 4308--4313, 2023.

\bibitem{pare2021multi}
Philip~E Par{\'e}, Ji~Liu, Carolyn~L Beck, Angelia Nedi{\'c}, and Tamer
  Ba{\c{s}}ar.
\newblock Multi-competitive viruses over time-varying networks with mutations
  and human awareness.
\newblock {\em Automatica}, 123:109330, 2021.

\bibitem{parino2024optimal}
Francesco Parino, Lorenzo Zino, and Alessandro Rizzo.
\newblock Optimal control of endemic epidemic diseases with behavioral
  response.
\newblock {\em IEEE Open Journal of Control Systems}, 3:483 -- 496, 2024.

\bibitem{prakash2012winner}
B~Aditya Prakash, Alex Beutel, Roni Rosenfeld, and Christos Faloutsos.
\newblock Winner takes all: competing viruses or ideas on fair-play networks.
\newblock In {\em International Conference on World Wide Web}, pages
  1037--1046, 2012.

\bibitem{rantzer2011distributed}
Anders Rantzer.
\newblock Distributed control of positive systems.
\newblock In {\em 50th IEEE Conference on Decision and Control and European
  Control Conference}, pages 6608--6611. IEEE, 2011.

\bibitem{reluga2010game}
Timothy~C Reluga.
\newblock Game theory of social distancing in response to an epidemic.
\newblock {\em PLoS Computational Biology}, 6(5):e1000793, 2010.

\bibitem{sanche2020novel}
Steven Sanche, Yen~Ting Lin, Chonggang Xu, Ethan Romero-Severson, Nicolas~W
  Hengartner, and Ruian Ke.
\newblock The novel coronavirus, 2019-n{C}o{V}, is highly contagious and more
  infectious than initially estimated.
\newblock {\em arXiv preprint arXiv:2002.03268}, 2020.

\bibitem{sandholm2010population}
William~H Sandholm.
\newblock {\em Population Games and Evolutionary Dynamics}.
\newblock MIT press, 2010.

\bibitem{sanjuan2016mechanisms}
Rafael Sanju{\'a}n and Pilar Domingo-Calap.
\newblock Mechanisms of viral mutation.
\newblock {\em Cellular and Molecular Life Sciences}, 73:4433--4448, 2016.

\bibitem{santos2015sufficient}
Augusto Santos, Jos{\'e}~MF Moura, and Jo{\~a}o~MF Xavier.
\newblock Sufficient condition for survival of the fittest in a bi-virus
  epidemics.
\newblock In {\em Asilomar Conference on Signals, Systems and Computers}, pages
  1323--1327, 2015.

\bibitem{satapathi2022epidemic}
Abhisek Satapathi, Narendra~Kumar Dhar, Ashish~R. Hota, and Vaibhav Srivastava.
\newblock Coupled evolutionary behavioral and disease dynamics under
  reinfection risk.
\newblock {\em IEEE Transactions on Control of Network Systems},
  11(2):795--807, 2024.

\bibitem{she2022networked}
Baike She, Ji~Liu, Shreyas Sundaram, and Philip~E Par{\'e}.
\newblock On a networked {SIS} epidemic model with cooperative and antagonistic
  opinion dynamics.
\newblock {\em IEEE Transactions on Control of Network Systems},
  9(3):1154--1165, 2022.

\bibitem{strogatz2024nonlinear}
Steven~H Strogatz.
\newblock {\em Nonlinear Dynamics and Chaos: With Applications to Physics,
  Biology, Chemistry, and Engineering}.
\newblock Chapman and Hall/CRC, 2024.

\bibitem{watkins2015optimal}
Nicholas~J Watkins, Cameron Nowzari, Victor~M Preciado, and George~J Pappas.
\newblock Optimal resource allocation for competing epidemics over arbitrary
  networks.
\newblock In {\em American Control Conference}, pages 1381--1386, 2015.

\bibitem{zhang2022networked}
Ciyuan Zhang, Sebin Gracy, Tamer Ba{\c{s}}ar, and Philip~E Par{\'e}.
\newblock A networked competitive multi-virus {SIR} model: Analysis and
  observability.
\newblock In {\em IFAC Conference on Networked Systems (NECSYS)}, pages 13--18,
  Z{\" u}rich, Switzerland, 2022.

\end{thebibliography}
                                 % bibliography (preferred). The
                                 % correct style is generated by
                                 % Elsevier at the time of printing.

\end{document}